\documentclass[pre,twocolumn, showpacs,amsmath,amssymb]{revtex4}
\usepackage{graphicx}
\usepackage{texdraw}

\usepackage{mathtext}

\newsavebox{\cl}
\sbox{\cl}{
\begin{texdraw}
\move (0 0) \fcir f:0.9 r:0.028 \larc r:0.028 sd:0 ed:360
\end{texdraw}}

\newsavebox{\cld}
\sbox{\cld}{
\begin{texdraw}
\move (0 0) \fcir f:1.0 r:0.028 \larc r:0.028 sd:0 ed:360
\end{texdraw}}

\newsavebox{\clu}
\sbox{\clu}{
\begin{texdraw}
\move (0 0) \fcir f:0.0 r:0.028 \larc r:0.028 sd:0 ed:360
\end{texdraw}}

\newsavebox{\alll}
\sbox{\alll}{
\unitlength=0.40mm%
\begin{picture}(20,10)
\put(2.9,0){\usebox{\cl}}
\put(2.9,5.1){\usebox{\cl}}
\put(8,0){\usebox{\cl}}
\end{picture}}

\newsavebox{\ddd}
\sbox{\ddd}{
\unitlength=0.40mm%
\begin{picture}(9,5)
\put(0.3,1.2){$\diagdown$}
\put(0.5,1.4){$\diagdown$}
\put(0.7,1.6){$\diagdown$}
\put(0,-0.7){\usebox{\cld}}
\put(0,4.4){\usebox{\cld}}
\put(5.1,-0.7){\usebox{\cld}}
\end{picture}}

\newsavebox{\udd}
\sbox{\udd}{
\unitlength=0.40mm%
\begin{picture}(9,5)
\put(0.3,1.2){$\diagdown$}
\put(0.5,1.4){$\diagdown$}
\put(0.7,1.6){$\diagdown$}
\put(0,-0.7){\usebox{\clu}}
\put(0,4.4){\usebox{\cld}}
\put(5.1,-0.7){\usebox{\cld}}
\end{picture}}

\newsavebox{\dud}
\sbox{\dud}{
\unitlength=0.40mm%
\begin{picture}(9,5)
\put(0.3,1.2){$\diagdown$}
\put(0.5,1.4){$\diagdown$}
\put(0.7,1.6){$\diagdown$}
\put(0,-0.7){\usebox{\cld}}
\put(0,4.4){\usebox{\clu}}
\put(5.1,-0.7){\usebox{\cld}}
\end{picture}}

\newsavebox{\uud}
\sbox{\uud}{
\unitlength=0.40mm%
\begin{picture}(9,5)
\put(0.3,1.2){$\diagdown$}
\put(0.5,1.4){$\diagdown$}
\put(0.7,1.6){$\diagdown$}
\put(0,-0.7){\usebox{\clu}}
\put(0,4.4){\usebox{\clu}}
\put(5.1,-0.7){\usebox{\cld}}
\end{picture}}

\newsavebox{\duu}
\sbox{\duu}{
\unitlength=0.40mm%
\begin{picture}(9,5)
\put(0.3,1.2){$\diagdown$}
\put(0.5,1.4){$\diagdown$}
\put(0.7,1.6){$\diagdown$}
\put(0,-0.7){\usebox{\cld}}
\put(0,4.4){\usebox{\clu}}
\put(5.1,-0.7){\usebox{\clu}}
\end{picture}}

\newsavebox{\uuu}
\sbox{\uuu}{
\unitlength=0.40mm%
\begin{picture}(9,5)
\put(0.3,1.2){$\diagdown$}
\put(0.5,1.4){$\diagdown$}
\put(0.7,1.6){$\diagdown$}
\put(0,-0.7){\usebox{\clu}}
\put(0,4.4){\usebox{\clu}}
\put(5.1,-0.7){\usebox{\clu}}
\end{picture}}

\newsavebox{\sdddd}
\sbox{\sdddd}{
\unitlength=0.40mm%
\begin{picture}(9,5)
\put(0.7,1.0){$\diagup$}
\put(0.5,1.0){$\diagup$}
\put(0.3,1.0){$\diagup$}
\put(0,-1.1){\usebox{\cld}}
\put(0,4.0){\usebox{\cld}}
\put(5.1,-1.1){\usebox{\cld}}
\put(5.1,4.0){\usebox{\cld}}
\end{picture}}

\newsavebox{\suddd}
\sbox{\suddd}{
\unitlength=0.40mm%
\begin{picture}(9,5)
\put(0.7,1.0){$\diagup$}
\put(0.5,1.0){$\diagup$}
\put(0.3,1.0){$\diagup$}
\put(0,-1.1){\usebox{\clu}}
\put(0,4.0){\usebox{\cld}}
\put(5.1,-1.1){\usebox{\cld}}
\put(5.1,4.0){\usebox{\cld}}
\end{picture}}

\newsavebox{\sdudd}
\sbox{\sdudd}{
\unitlength=0.40mm%
\begin{picture}(9,5)
\put(0.7,1.0){$\diagup$}
\put(0.5,1.0){$\diagup$}
\put(0.3,1.0){$\diagup$}
\put(0,-1.1){\usebox{\cld}}
\put(0,4.0){\usebox{\clu}}
\put(5.1,-1.1){\usebox{\cld}}
\put(5.1,4.0){\usebox{\cld}}
\end{picture}}

\newsavebox{\suudd}
\sbox{\suudd}{
\unitlength=0.40mm%
\begin{picture}(9,5)
\put(0.7,1.0){$\diagup$}
\put(0.5,1.0){$\diagup$}
\put(0.3,1.0){$\diagup$}
\put(0,-1.1){\usebox{\clu}}
\put(0,4.0){\usebox{\clu}}
\put(5.1,-1.1){\usebox{\cld}}
\put(5.1,4.0){\usebox{\cld}}
\end{picture}}

\newsavebox{\suddu}
\sbox{\suddu}{
\unitlength=0.40mm%
\begin{picture}(9,5)
\put(0.7,1.0){$\diagup$}
\put(0.5,1.0){$\diagup$}
\put(0.3,1.0){$\diagup$}
\put(0,-1.1){\usebox{\clu}}
\put(0,4.0){\usebox{\cld}}
\put(5.1,-1.1){\usebox{\cld}}
\put(5.1,4.0){\usebox{\clu}}
\end{picture}}

\newsavebox{\sduud}
\sbox{\sduud}{
\unitlength=0.40mm%
\begin{picture}(9,5)
\put(0.7,1.0){$\diagup$}
\put(0.5,1.0){$\diagup$}
\put(0.3,1.0){$\diagup$}
\put(0,-1.1){\usebox{\cld}}
\put(0,4.0){\usebox{\clu}}
\put(5.1,-1.1){\usebox{\clu}}
\put(5.1,4.0){\usebox{\cld}}
\end{picture}}

\newsavebox{\sduuu}
\sbox{\sduuu}{
\unitlength=0.40mm%
\begin{picture}(9,5)
\put(0.7,1.0){$\diagup$}
\put(0.5,1.0){$\diagup$}
\put(0.3,1.0){$\diagup$}
\put(0,-1.1){\usebox{\cld}}
\put(0,4.0){\usebox{\clu}}
\put(5.1,-1.1){\usebox{\clu}}
\put(5.1,4.0){\usebox{\clu}}
\end{picture}}

\newsavebox{\suduu}
\sbox{\suduu}{
\unitlength=0.40mm%
\begin{picture}(9,5)
\put(0.7,1.0){$\diagup$}
\put(0.5,1.0){$\diagup$}
\put(0.3,1.0){$\diagup$}
\put(0,-1.1){\usebox{\clu}}
\put(0,4.0){\usebox{\cld}}
\put(5.1,-1.1){\usebox{\clu}}
\put(5.1,4.0){\usebox{\clu}}
\end{picture}}

\newsavebox{\suuuu}
\sbox{\suuuu}{
\unitlength=0.40mm%
\begin{picture}(9,5)
\put(0.7,1.0){$\diagup$}
\put(0.5,1.0){$\diagup$}
\put(0.3,1.0){$\diagup$}
\put(0,-1.1){\usebox{\clu}}
\put(0,4.0){\usebox{\clu}}
\put(5.1,-1.1){\usebox{\clu}}
\put(5.1,4.0){\usebox{\clu}}
\end{picture}}

\newsavebox{\dddd}
\sbox{\dddd}{
\unitlength=0.40mm%
\begin{picture}(9,5)
\put(0,-1.1){\usebox{\cld}}
\put(0,4.0){\usebox{\cld}}
\put(5.1,-1.1){\usebox{\cld}}
\put(5.1,4.0){\usebox{\cld}}
\end{picture}}

\newsavebox{\uddd}
\sbox{\uddd}{
\unitlength=0.40mm%
\begin{picture}(9,5)
\put(0,-1.1){\usebox{\clu}}
\put(0,4.0){\usebox{\cld}}
\put(5.1,-1.1){\usebox{\cld}}
\put(5.1,4.0){\usebox{\cld}}
\end{picture}}

\newsavebox{\uudd}
\sbox{\uudd}{
\unitlength=0.40mm%
\begin{picture}(9,5)
\put(0,-1.1){\usebox{\clu}}
\put(0,4.0){\usebox{\clu}}
\put(5.1,-1.1){\usebox{\cld}}
\put(5.1,4.0){\usebox{\cld}}
\end{picture}}

\newsavebox{\uddu}
\sbox{\uddu}{
\unitlength=0.40mm%
\begin{picture}(9,5)
\put(0,-1.1){\usebox{\clu}}
\put(0,4.0){\usebox{\cld}}
\put(5.1,-1.1){\usebox{\cld}}
\put(5.1,4.0){\usebox{\clu}}
\end{picture}}

\newsavebox{\duuu}
\sbox{\duuu}{
\unitlength=0.40mm%
\begin{picture}(9,5)
\put(0,-1.1){\usebox{\cld}}
\put(0,4.0){\usebox{\clu}}
\put(5.1,-1.1){\usebox{\clu}}
\put(5.1,4.0){\usebox{\clu}}
\end{picture}}

\newsavebox{\uuuu}
\sbox{\uuuu}{
\unitlength=0.40mm%
\begin{picture}(9,5)
\put(0,-1.1){\usebox{\clu}}
\put(0,4.0){\usebox{\clu}}
\put(5.1,-1.1){\usebox{\clu}}
\put(5.1,4.0){\usebox{\clu}}
\end{picture}}

\begin{document}

\title{Ground-state structures in Ising magnets on the Shastry-Sutherland
lattice with long-range interactions and fractional magnetization
plateaus in TmB${}_4$}

\author{Yu. I. Dublenych}
\affiliation{Institute for Condensed Matter Physics, National
Academy of Sciences of Ukraine, 1 Svientsitskii Street, 79011
Lviv, Ukraine}
\date{\today}

\begin{abstract}{A method for the study of the ground states of
lattice-gas models or equivalent spin models with extended-range
interactions is developed. It is shown that effect of longer-range
interactions can be studied in terms of the solution of the
ground-state problem for a model with short-range interactions.
The method is applied to explain the emergence of fractional
magnetization plateaus in TmB${}_4$ that is regarded as a strong
Ising magnet on the Shastry-Sutherland lattice.}
\end{abstract}

\maketitle

\section{Introduction}

Exact determination of the ground-state structures for complex
lattice-gas models or equivalent spin models still remains an open
problem despite considerable effort made for more than half a
century \cite{bib1, bib2}. Many methods, both analytical and
numerical, have been proposed \cite{bib3, bib4, bib5, bib6},
however, an universal effective algorithm has not been found as
yet. We have elaborated a new method for the study of ground
states for such models and successfully applied it to some
interesting physical problems \cite{bib7, bib8, bib9, bib10,
bib11, bib12}. In the present paper, we develop the method in
order to show how to treat (at least partially) the effect of
longer-range interactions in terms of the solution of the
ground-state problem for a model with short-range interactions. We
demonstrate this by considering a system of Ising spins on the
Shastry-Sutherland (SS) lattice [see Fig.~1(a)] in the presence of
an external magnetic field. The ground-state magnetic structures
of this system are interesting because they are associated with
the emergence of fractional magnetization plateaus in some
rare-earth-metal tetraborides, particularly in TmB${}_4$ regarded
as a strong Ising magnet \cite{bib14,bib15}.

This compound consists of weakly coupled layers of magnetic ions
Tm$^{3+}$ arranged on a lattice that is topologically equivalent
to the SS one. Experiments show that, in addition to a large
1/2-magnetization plateau, TmB${}_4$ exhibits a sequence of narrow
fractional magnetization plateaus at 1/6, 1/7, up to 1/12 of the
saturation magnetization for temperatures below 4 K, with the
magnetic field being normal to SS planes \cite{bib14,bib15}. In
spite of considerable effort made to find the origin of these
plateaus, only the 1/2-plateau has been reliably obtained in some
theoretical works. Hence, the question remains open.

In Ref.~\onlinecite{bib12}, we found a complete solution of the
ground-state problem for the Ising model on an extended SS lattice
[Fig.~1(a)], i.e., with an interaction along the diagonals of
``empty'' squares (without SS bonds) in addition to the
interactions along the edges of squares and the SS diagonals. In
this model, the existence of a 1/2 plateau was proved. We have
also shown that magnetic structures that can generate other
fractional plateaus in TmB${}_4$ are the ground-state ones at some
boundaries of the full-dimensional ground-state regions of the
four-dimensional parameter space of the model. At all the
boundaries, a degeneracy exists. This degeneracy is at least
twofold (in the case when only two nondegenerate phases exist at a
boundary between these phases). But usually, the degeneracy is
infinite and uncountable and often even macroscopic (i.e., leading
to residual entropy).

Here, we determine the range of interactions which lift (at least
partially) the degeneracy at three-dimensional boundaries of full
(four)-dimensional ground-state regions as well as emerging
full-dimensional (in the extended parameter space) phases that
give rise to new magnetization plateaus. We do not consider all
the three-dimensional boundaries but only those which can be
associated with the emergence of fractional magnetization plateaus
in rare-earth-metal tetraborides, particularly in TmB${}_4$.

\begin{figure}[b]
\begin{center}
\includegraphics[scale = 1.35]{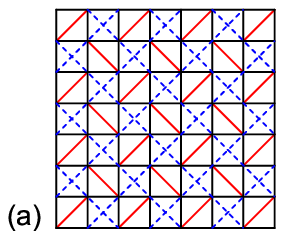}
\hspace{0.5cm}
\includegraphics[scale = 1.06]{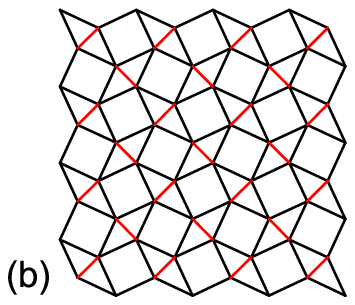}
\caption{(Color online) (a) Extended Shastry-Sutherland lattice
and (b) the lattice formed by magnetic Cu$^{2+}$ ions in
SrCu${}_2$(BO${}_3$)$_2$.}
\label{fig1}
\end{center}
\end{figure}

Although we consider a specific problem, the method developed here
is general and may be applied to many other problems. The method
is based on the notion of fractional contents of cluster
configurations in the structures generated by these configurations
and on some linear relations between these contents \cite{bib13}.

To enumerate pairwise interactions, we use the lattice shown in
Fig.~1(b). The coordination circles for this lattice are shown in
Fig.~2. We designate the $i$th-neighbor interaction on this
lattice by $J_i$ except for the first- and second-neighbor
interactions which are denoted by $\tilde J_1$ and $\tilde J_2$,
respectively, in order to avoid the confusion with the notations
introduced in Refs.~\onlinecite{bib11} and \onlinecite{bib12}.

\begin{figure}[h]
\begin{center}
\includegraphics[scale = 2.25]{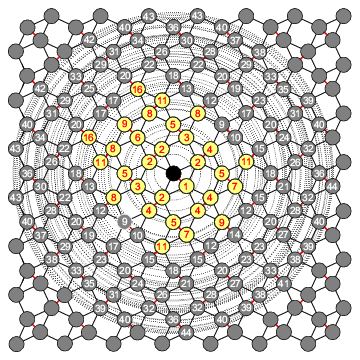}

\vspace{0.25cm}

\includegraphics[scale = 1.75]{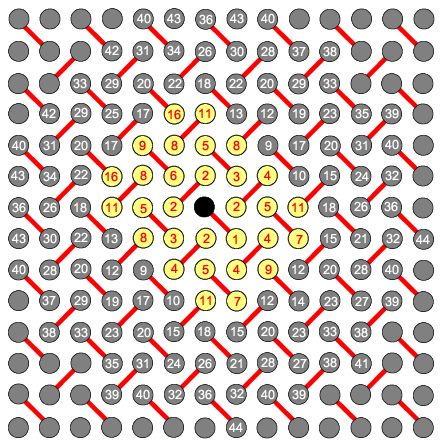}
\caption{Top: Coordination circles and respective neighbors of the
site depicted in black on the lattice that is topologically
equivalent to the SS lattice. Bottom: Similar neighbors on the SS
lattice. The sites contained in a ``windmill'' cluster (see
Fig.~4) together with the central (black) site are depicted in
yellow.}
\label{fig2}
\end{center}
\end{figure}

We construct structures on the lattice shown in Fig.~1(a), but
name the clusters after their shapes on the lattice shown in
Fig.~1(b). The clusters which we use here are depicted in Fig.~3.
Similar cluster configurations are enumerated differently for
different boundaries between full-dimensional regions.

\begin{figure}[]
\begin{center}
\includegraphics[scale = 1.5]{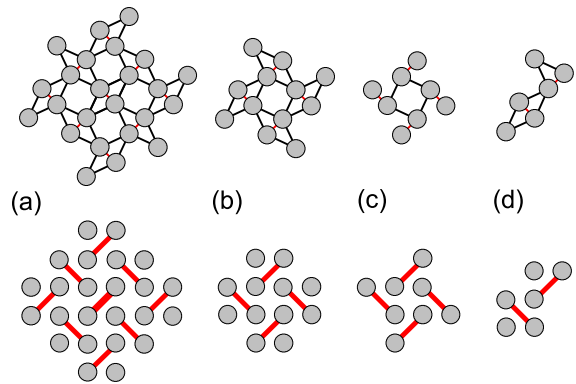}
\caption{Clusters considered here: (a) ``turtle'', (b)
``windmill'', (c) ``screw'', (d) ``seahorse''. At the top and at
the bottom, the clusters are depicted on the lattice shown in
Fig.~1(b) and in Fig.~1(a), respectively.}
\label{fig3}
\end{center}
\end{figure}

\section{Full-dimensional ground-state structures emerging from
the boundary between the N\'{e}el phase and the 1/3-plateau phase}

In our previous papers, we have shown that, at the boundary
between the N\'{e}el phase (phase 3) and the 1/3-plateau phase
(phase 4) (see Fig.~4), the ground-state structures consist of the
following configurations of the triangular cluster with SS bond as
hypotenuse: \usebox{\udd}, \usebox{\uud}, and \usebox{\duu}
\cite{bib11}, or the equivalent set of square configurations,
i.e., \usebox{\suddu}, \usebox{\sduud}, \usebox{\sduuu};
\usebox{\uddu}, and \usebox{\duuu} \cite{bib12} [herein solid and
open circles denote spin up ($\sigma = +1$) and down ($\sigma =
-1$), respectively]. This means that any triangular or square
cluster in any ground-state structure at this boundary should have
one of the listed configurations. It is easy to see that these
structures represent antiferromagnetic stripes of various widths
(domains of the N\'{e}el phase with even numbers of
antiferromagnetic chains) separated by ferromagnetic chains [see
Fig.~4(c)]. Longer-range interactions lift the degeneracy
partially and lead to the emergence of new full-dimensional
phases, that is, to the appearance of new fractional magnetization
plateaus.

To study the effect of longer-range interactions, first, let us
consider a windmill-shaped cluster. Its configurations generating
all the structures at the boundary between phases 3 and 4 are
shown in Fig.~5. It means that, in any ground-state structure at
this boundary, any cluster of this type on the lattice has one of
these configurations (or chiral ones). The structure shown in
Fig.~6 is similar to that in Fig.~4(c) but with the number of the
``windmill'' configuration indicated in the center of each
``empty'' square.

\begin{figure}[h]
\begin{center}
\includegraphics[scale = 1.0]{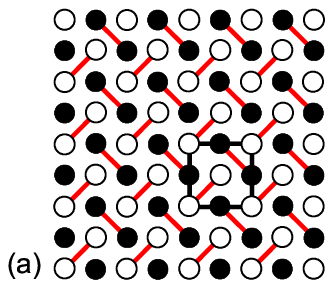}
\hspace{0.5cm}
\includegraphics[scale = 1.0]{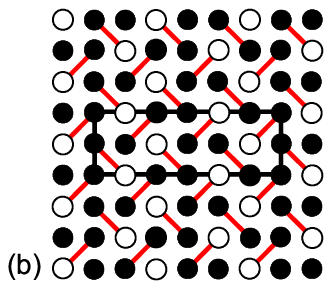}

\vspace{0.5cm}

\includegraphics[scale = 1.0]{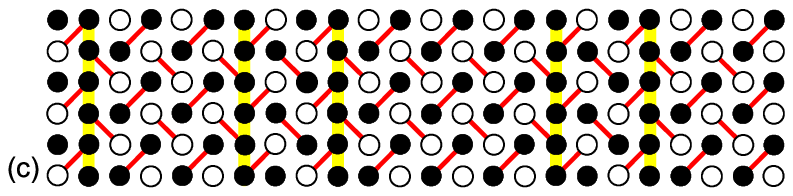}
\caption{(a) N\'{e}el structure, (b) 1/3-plateau structure, and
(c) a general disordered structure at the boundary of the relative
phases (the yellow background indicates ferromagnetic chains).}
\label{fig4}
\end{center}
\end{figure}

\begin{figure}[h]
\begin{center}
\includegraphics[scale = 1.2]{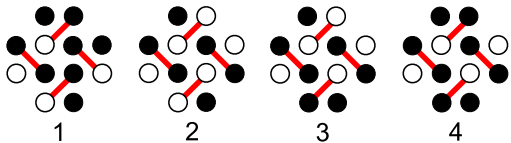}
\caption{Configurations of the ``windmill'' cluster for the
structures at the boundary between the N\'{e}el phase and the
1/3-plateau phase.}
\label{fig5}
\end{center}
\end{figure}

\begin{figure}[]
\begin{center}
\includegraphics[scale = 1.0]{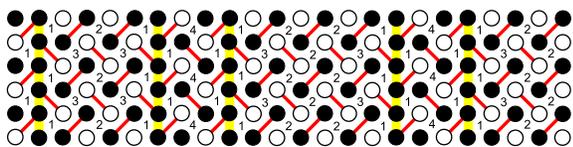}
\caption{The similar structure as in Fig.~4(c), but with the
number of the ``windmill'' configuration indicated in the center
of each ``empty'' square.}
\label{fig6}
\end{center}
\end{figure}

\begin{figure}[]
\begin{center}
\includegraphics[scale = 2.0]{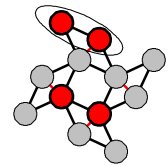}
\caption{A subcluster of the ``windmill'' cluster and its two
nonequivalent positions in the cluster: central and lateral
(enveloped by an ellipse).}
\label{fig7}
\end{center}
\end{figure}

We refer to the relative quantity of a configuration of a cluster
in a structure as a fractional content of the configuration in the
structure. Let $k_1$, $k_2$, $k_3$, and $k_4$ be the fractional
contents of the configurations shown in Fig.~5 in the structures
at the boundary between phases 3 and 4. In addition to the trivial
relation between $k_i$ (the normalization condition),
\begin{equation}
k_1 + k_2 + k_3 + k_4 = 1,
\label{eq2}
\end{equation}
there is one more linear relation between these quantities (see
\cite{bib13}). To find it, let us consider a subcluster of the
``windmill'' cluster that has at least two nonequivalent positions
in this cluster. Let it be the two-site subcluster shown in
Fig.~7. It can occupy even four nonequivalent positions in the
``windmill'' cluster. Consider two of these: the central one
(position 1) and the lateral one (position 2, enveloped by an
ellipse in Fig.~7). In each position, the subcluster enters only
one ``windmill'' cluster on the lattice: $c_1 = c_2 = 1$. Let us
consider the ``two spins up'' configuration of the subcluster and
calculate the number of such configurations in each position for
each configuration of the cluster: $n_{11} = 2$, $n_{12} = 0$,
$n_{13} = 0$, $n_{14} = 0$, $n_{21} = 1$, $n_{22} = 0$, $n_{23} =
1$, and $n_{24} = 2$. Using the general relation
\begin{equation}
\sum_l \frac{k_l n_{1l}}{c_1} = \sum_l\frac{k_l n_{2l}}{c_2},
\label{eq3}
\end{equation}
where $l$ is the number of the cluster configuration, we have
\begin{equation}
k_1 - k_3 - 2k_4 = 0.
\label{eq4}
\end{equation}
Other configurations of the subcluster or other subclusters yield
the same relation. Hence, two of the four quantities $k_1$, $k_2$,
$k_3$, and $k_4$ are independent, for instance, $k_1$ and $k_2$,
and the other two can be linearly expressed in terms of these,
i.e.,
\begin{equation}
k_3 = -3k_1 - 2k_2 + 2,~~ k_4 =  2k_1 + k_2 - 1.
\label{eq5}
\end{equation}

Bearing in mind that each site on the lattice belongs to six
``windmill'' clusters, we can find the magnetization per site for
each structure generated by the set of cluster configurations
under consideration, i.e.,
\begin{equation}
m = \frac16 (2k_1 + k_3 + 2k_4) = \frac{k_1}{2}.
\label{eq6}
\end{equation}
Moreover, the magnetization per site is given by the relative
number of ferromagnetic chains in the structure. Hence, this
number is equal to $\frac{k_1}{2}$. In order $k_2$ to be maximum
(minimum) for fixed $k_1$ (that is, for fixed number of
ferromagnetic chains), the number of narrowest stripes (and hence,
of configurations 4) should be maximum (minimum). This is clear
from Fig.~6. Thus the region of variation for the quantities $k_1$
and $k_2$ is the triangle $ABC$ shown in Fig.~8. The vertex with
the coordinates $k_1 = 0$, $k_2 = 1$ ($k_3 = k_4 = 0$) corresponds
to the N\'{e}el structure, and the vertex $k_1 = 2/3$, $k_2 = 0$
($k_3 = 0$, $k_4 = 1/3$) corresponds to the 1/3-plateau structure.
The vertex with the coordinates $k_1 = 2/5$, $k_2 = 1/5$ ($k_3 =
2/5$, $k_4 = 0$) corresponds to the only structure shown in
Fig.~9.

\begin{figure}[]
\begin{center}
\includegraphics[scale = 0.75]{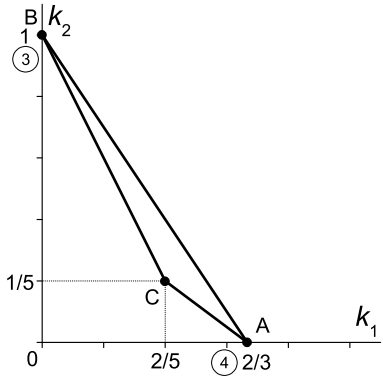}
\caption{Region of variation for the quantities $k_1$ and $k_2$.
Vertexes $B(0,~1)$, $A(2/3,~0)$, and $C(2/5,~1/5)$ of the triangle
$ABC$ correspond to the N\'{e}el structure, the 1/3-plateau
structure, and the 1/5-plateau structure, respectively.}
\label{fig8}
\end{center}
\end{figure}

\begin{figure}[h]
\begin{center}
\includegraphics[scale = 1.0]{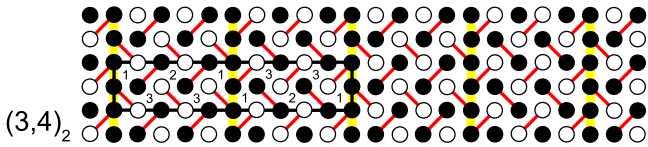}
\caption{1/5-plateau structure that emerges from the boundary
between phases 3 and 4, if the interaction range reaches the
seventh neighbors ($k_1 = k_3 = 2/5$, $k_2 = 1/5$).}
\label{fig9}
\end{center}
\end{figure}

The fact that the region of variation for the quantities $k_1$ and
$k_2$ is the triangle $ABC$, can be proved in other way, using
relations (4). From the first of these we have the inequality
\begin{equation}
3k_1 + 2k_2 - 2 = -k_3 \leqslant 0,
\label{eq7}
\end{equation}
which, being regarded as an equation, represents an equation of
the straight line $AB$ ($k_3 = 0$). The second relation
\begin{equation}
2k_1 + k_2 - 1 = k_4
\label{eq8}
\end{equation}
represents an equation of the straight line $BC$ for which $k_4 =
0$. For $k_4 = 0$, however, the minimum value of $k_2$ is equal to
1/5 rather than 0. This follows from the relation
\begin{equation}
3k_1 + 4k_2 - 2 = 2k_2 - k_3 \geqslant  0
\label{eq9}
\end{equation}
and the inequality $2k_2 \geqslant k_3$. Thus, we obtain the point
$C(2/5, 1/5)$ that corresponds to the structure shown in Fig.~9.

If all the interactions (pairwise as well as many-spin) occur
within the ``windmill'' cluster,  then the energy of a
ground-state structure is a linear function of $k_1$ and $k_2$,
and therefore (for fixed values of interactions and external
field) it can reach the minimum value only at the boundary of the
triangle shown in Fig.~8. Therefore, only the vertexes of the
triangle correspond to the full-dimensional structures. Hence, the
interactions within the ``windmill'' cluster can generate only one
full-dimensional structure from the boundary between the N\'{e}el
phase and the 1/3-plateau phase. This structure is shown in
Fig.~9; it gives rise to the plateau with the magnetization 1/5.
To determine the range of interactions which can generate this
structure, let us find the contribution of various pairwise
interactions (within the ``windmill'' cluster) into the energy
density (i.e., energy per site). We have
\begin{eqnarray}
&&e_1 = \frac14 (-k_1 + 2k_2 + k_3)\tilde J_1 = (-k_1 + \frac12)\tilde J_1,\nonumber\\
&&e_2 = -\frac12 (k_1 + 4k_2 + 3k_3 + 2k_4)\tilde J_2 = (2k_1 - 2)\tilde J_2,\nonumber\\
&&e_3 = (k_2 + k_3 + k_4)J_3 = (-k_1 + 1)J_3,\nonumber\\
&&e_4 = (k_1 - 2k_2 - k_3)J_4 = (4k_1 - 2)J_4,\nonumber\\
&&e_5 = (k_1 + 2k_2 + k_3)J_5 = (-2k_1 + 2)J_5,\nonumber\\
&&e_6 = \frac12 J_6,\nonumber\\
&&e_7 = (-k_1 + k_2 + k_4)J_7 = (k_1 + 2k_2 - 1)J_7,\nonumber\\
&&e_8 = -2(k_2 + k_3 + k_4)J_8 = (2k_1 - 2)J_8,\nonumber\\
&&e_{9a} = \frac12(-k_1 + 2k_2 +k_3)J_9 = (-2k_1 + 1)J_9,\nonumber\\
&&e_{11} = (k_1 - 2k_2 - k_3)J_{11} = (4k_1 - 2)J_{11},\nonumber\\
&&e_{16} = (k_2 + k_3 + k_4)J_{16} = (-k_1 + 1)J_{16}.
\label{eq10}
\end{eqnarray}
Since only a half of the number of ninth neighbors enter the
``windmill'' cluster (see Fig.~2), we  denote their contribution
to the energy density by $e_{9a}$.

Thus, contributions $e_i$ to the energy density of all the
pairwise interactions within the ``windmill'' cluster except for
the seventh neighbors depend on the magnetization of the structure
only, that is, on $k_1$. However, $e_7$ depends also on $k_2$.
This means that only the pairwise interaction of the seventh
neighbors or many-spin interactions which include the seventh
neighbors can lift the degeneracy at the boundary between the
N\'{e}el phase and the 1/3-plateau phase giving rise to a new
full-dimensional phase. This is the phase $(3,4)_2$ (see
Ref.~\onlinecite{bib11} about the notations) with the
magnetization 1/5 (Fig~9).

\begin{figure}[h]
\begin{center}
\includegraphics[scale = 1.0]{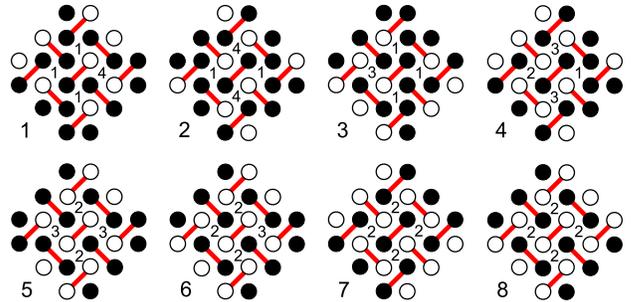}
\caption{Configurations of the ``turtle'' cluster which generate
all the ground-state structures at the boundary between the
N\'{e}el phase and the 1/3-plateau phase. Each ``empty'' square
contains the number of the ``windmill'' configuration with the
center in this square (see Fig.~5).}
\label{fig10}
\end{center}
\end{figure}

In order to take into account interactions beyond the ``windmill''
cluster, we consider a bigger cluster whose shape resembles a
turtle. The ground-state configurations of this cluster for the
boundary between the N\'{e}el phase and the 1/3-plateau phase are
shown in Fig.~10.

\begin{figure}[]
\begin{center}
\includegraphics[scale = 1.75]{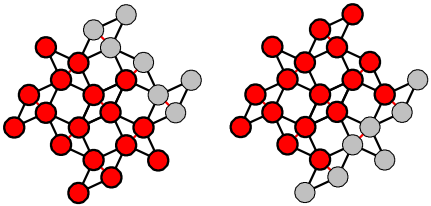}
\caption{Maximum subcluster which can occupy two nonequivalent
positions in the ``turtle'' cluster.}
\label{fig11}
\end{center}
\end{figure}

Considering the configurations of the maximum subcluster among
those which can occupy two nonequivalent positions in the
``turtle'' cluster (Fig.~11), we find the relations between the
fractional contents $l_i$ of these configurations in the
ground-state structures to be given by
\begin{eqnarray}
&&l_1 - 2l_2 = 0,\nonumber\\
&&l_3 - l_4 = 0,\nonumber\\
&&l_4 - 2l_5 - l_6 = 0,\nonumber\\
&&l_6 - 2l_7 + 2l_8 = 0.
\label{eq11}
\end{eqnarray}
Thus, only three of the eight normalized quantities $l_i$ are
independent. Let it be $l_1$, $l_3$, and $l_5$. The rest of the
quantities $l_i$ can be expressed in terms of these, i.e.,
\begin{eqnarray}
&&l_2 = \frac{1}{2}l_1,\nonumber\\
&&l_4 = l_3,\nonumber\\
&&l_6 = l_3 - 2l_5,\nonumber\\
&&l_7 = -\frac{3}{4}l_1 - \frac{5}{4}l_3 + \frac12,\nonumber\\
&&l_8 = -\frac{3}{4}l_1 - \frac{7}{4}l_3 + l_5 + \frac12.
\label{eq12}
\end{eqnarray}

The quantities $k_i$ can be expressed in terms of $l_1$ and $l_3$
as well. To do this we just have to calculate the number of
configurations of the ``windmill'' subcluster in the
configurations of the ``turtle'' cluster. Thus we have
\begin{eqnarray}
&&k_1 = l_1 + l_3,~~ k_2 = -\frac32 l_1 -2l_3 + 1,\nonumber\\
&&k_3 = l_3,~~ k_4 = \frac12 l_1.
\label{eq13}
\end{eqnarray}
As follows from these relations,
\begin{equation}
l_1 = 4k_1 + 2k_2 - 2,~~ l_3 = -3k_1 - 2k_2 + 2.
\label{eq14}
\end{equation}

Hence, in addition to the two independent quantities $k_1$ and
$k_2$, we have one more independent quantity, $l_5$. The
quantities $k_1$ and $k_2$ determine the number of stripes with
two antiferromagnetic chains (we denote this stripe by $s_1$), and
$l_5$ determines the number of stripes with four antiferromagnetic
chains (we denote this stripe by $s_2$). The region of variation
for $k_1$, $k_2$, and $l_5$ is shown in Fig.~12. This is the
pyramid $ABCD$. The points of the face $BCD$ correspond to the
structures without stripes $s_1$. The structures which correspond
to the edge $BD$ do not contain the stripes $s_1$ and $s_2$; in
the structures corresponding to the edge $BC$, the number of
stripes $s_2$ is maximum. The face $ABC$ corresponds to the
structures with maximum numbers of stripes $s_2$. The projection
of this pyramid on the $(k_1, k_2)$ plane is similar to the
triangle shown in Fig.~8.

\begin{figure}[]
\begin{center}
\includegraphics[scale = 0.75]{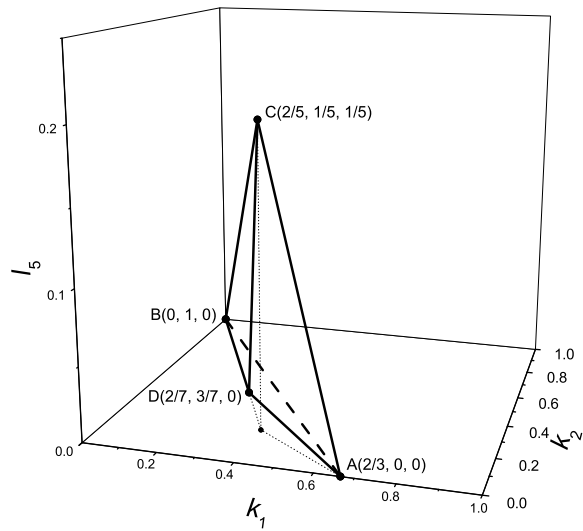}
\caption{Region of variation for the quantities $k_1$, $k_2$, and
$l_5$. Each vertex corresponds to a full-dimensional structure.}
\label{fig12}
\end{center}
\end{figure}

Let us prove rigorously that the tetrahedron $ABCD$ is indeed the
region of variation for the quantities $k_1$, $k_2$, and $l_5$. It
is determined by four inequalities given by
\begin{eqnarray}
&&ABC: 3k_1 + 2k_2 - 2 + 2l_5 \leqslant 0,\nonumber\\
&&ACD: 9k_1 + 8k_2 - 6 + 4l_5 \geqslant 0,\nonumber\\
&&BCD: 2k_1 + k_2 - 1 \geqslant 0,\nonumber\\
&&ABD: l_5 \geqslant 0.
\label{eq15}
\end{eqnarray}

The first one follows from relations $l_3 - 2l_5 = l_6$ and $l_3 =
-3k_1 - 2k_2 + 2$ [the third relation from (9) and the second one
from (12)]. Hence, for the face $ABC$, the value of $l_6$ is equal
to zero. The second inequality follows from the last relation of
(10) and relations (12). For the face $ACD$, the value of $l_8$ is
equal to zero. The inequality for the face $BCD$ is the same as
for the straight line $BC$ for which $k_4 = 0$ (i.e., the
structures on this face have no narrow stripes $s_1$).

\begin{figure}[]
\begin{center}
\includegraphics[scale = 1.0]{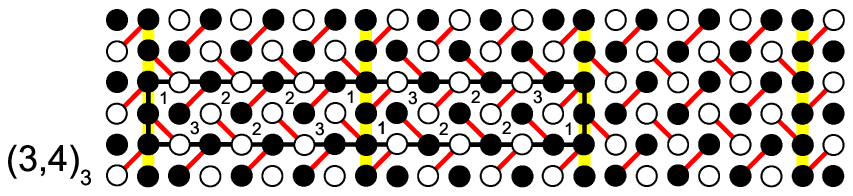}
\caption{Structure with magnetization 1/7 that emerges from the
boundary between phases 3 and 4 if the range of interaction
reaches the 21th neighbors (five square lattice constants). For
this structure $k_1 = \frac27$, $k_2 = \frac37$, and $l_5 = 0$.}
\label{fig13}
\end{center}
\end{figure}

\begin{figure}[]
\begin{center}
\includegraphics[scale = 1.0]{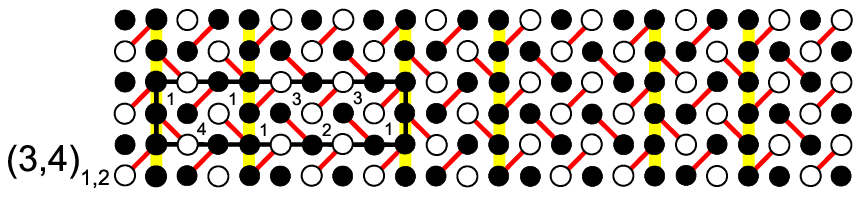}
\caption{Structure with magnetization 1/4 that emerges from the
boundary between phases 3 and 4 if the range of interaction
reaches the 32th neighbors (six square lattice constants). For
this structure $k_1 = \frac12$, $k_2 = \frac18$, and $l_5 =
\frac18$.}
\label{fig14}
\end{center}
\end{figure}

\begin{figure}[]
\begin{center}
\includegraphics[scale = 2.25]{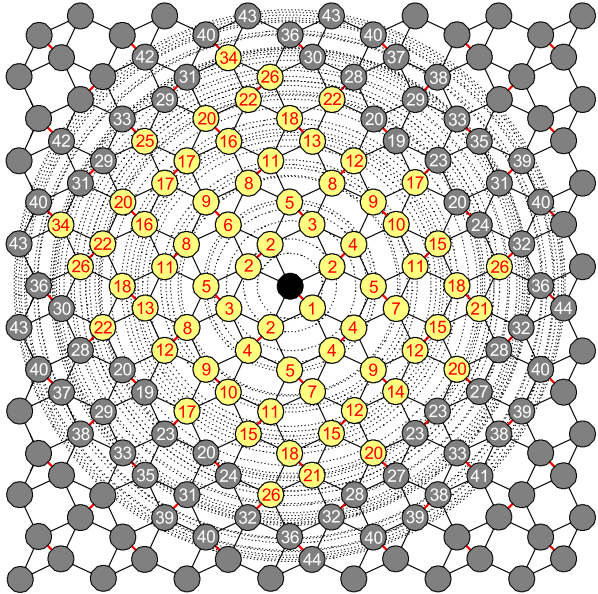}
\caption{Pairwise interactions on the SS lattice which can be
taken into account by considering the ``turtle'' cluster (see
Fig.~4). The corresponding neighbors of the central site (the
black circle) are depicted in yellow.}
\label{fig15}
\end{center}
\end{figure}

Thus, interactions within the ``turtle'' cluster which are absent
in the ``windmill'' cluster can generate only one additional
full-dimensional structure, the structure with  $k_1 = 2/7$, $k_2
= 3/7$, and $l_5 = 0$. It is shown in Fig.~13. The magnetization
of this structure is equal to 1/7.

To determinate the range of interactions which can give rise to
this structure, let us find the contribution to the energy density
of various pairwise interactions within the ``turtle'' cluster
which (except for 9th, 11th, and 16th neighbors) are absent in the
``windmill'' cluster (see Fig.~15). It should be noted that only
two sites belong to one ``turtle'' cluster on the lattice. Thus,
we have
\begin{eqnarray}
&&e_9 = 2(-l_1 + l_2 - l_3 + l_4 + l_5 + l_6 + l_7 + l_8)J_9\nonumber\\
&&~~~~= (-4k_1 + 2)J_9,\nonumber\\
&&e_{10} = (l_2 + l_5 + l_6 + l_7 + l_8)J_{10}\nonumber\\
&&~~~~~= (-l_1 - 2l_3 + 1)J_{10} = (2k_1 + 2k_2 - 1)J_{10},\nonumber\\%(-k_1 - k_3 + 1)
&&e_{11} = 2(l_2 - l_5 - l_6 - l_7 - l_8)J_{11} = (4k_1 - 2)J_{11},\nonumber\\
&&e_{12} = 2(l_1 + l_2 + l_3 - l_5 - l_6 - l_7 - l_8)J_{12}\nonumber\\
&&~~~~~= (6k_1 - 2)J_{12},\nonumber\\
&&e_{13} = (-l_1 + l_2 - l_3 + l_5 + l_6 + l_7 + l_8)J_{13}\nonumber\\
&&~~~~~= (-2k_1 + 1)J_{13},\nonumber\\
&&e_{14} = \frac12(-l_1 + l_2 - l_3 - l_4 + l_5 + l_6 + l_7 + l_8)J_{14}\nonumber\\
&&~~~~~= (-l_1 - 2l_3 + \frac12)J_{14} = (2k_1 + 2k_2 -\frac32)J_{14},\nonumber\\%(-k_1 - k_3 + \frac12)
&&e_{15} = (-2l_2 + l_3 - l_6 - 2l_7 - 2l_8)J_{15}\nonumber\\
&&~~~~~= (2l_1 + 6l_3 - 2)J_{15} = (-10k_1 - 8k_2 + 6)J_{15},\nonumber\\%(2k_1 + 4k_3 - 2)
&&e_{16} = (l_2 + l_4 + l_5 + l_6 + l_7 + l_8)J_{16}\nonumber\\
&&~~~~~= (-k_1 + 1)J_{16},\nonumber\\
&&e_{17} = (l_1 + l_3 - l_4 - 2l_5 - 2l_6 - 2l_7 - 2l_8)J_{17}\nonumber\\
&&~~~~~= (4k_1 - 2)J_{17},\nonumber\\
&&e_{18} = (l_1 + l_4 + l_6 + 2l_7 + 2l_8)J_{18}\nonumber\\
&&~~~~~= (-2l_1 - 4l_3 + 2)J_{18} = (4k_1 + 4k_2 - 2)J_{18},\nonumber\\%(-2k_1 - 2k_3 + 2)
&&e_{20a} = \frac{1}{4}(-4l_1 - 6l_3 + 4l_5 + 6l_6 + 8l_7 + 8l_8)J_{20}\nonumber\\
&&~~~~~= (-4l_1 - 6l_3 + 2)J_{20} = (2k_1 + 4k_2 - 2)J_{20},\nonumber\\%(-5k_1 - k_3 + 2)
&&e_{21} = (l_2 - l_3 + l_5 + l_7 + l_8)J_{21}\nonumber\\
&&~~~~~= (-l_1 - 4l_3 + 2l_5 + 1)J_{21}\nonumber\\
&&~~~~~= (8k_1 + 6k_2 + 2l_5 - 5)J_{21},\nonumber\\%(-k_1 - 3k_3 + 2l_5 + 1)
&&e_{22} = (l_1 + l_3 - l_4 - 2l_5 - 2l_6 - 2l_7 - 2l_8)J_{22}\nonumber\\
&&~~~~~= (4l_1 + 4l_3 - 2)J_{22} = (4k_1 - 2)J_{22},\nonumber\\%(3k_1 + k_3 - 2)
&&e_{25} = \frac{1}{6}(-l_1 - l_2 - l_3 + l_4 + 3l_5 + 3l_6 + 3l_7 + 3l_8)J_{25}\nonumber\\
&&~~~~~= \frac{1}{6}(-6l_1 - 6l_3 + 3)J_{25} = (-k_1 + \frac12)J_{25},\nonumber\\%(-5k_1 + 2k_3 + 2)J_{25}
&&e_{26} = \frac12(-4l_2 + 2l_3 - 2l_6 - 4l_7 - 4l_8)J_{26}\nonumber\\
&&~~~~~= (2l_1 + 6l_3 - 2)J_{26} = (-10k_1 - 8k_2 + 6)J_{26},\nonumber\\%(2k_1 + 5k_3 - 2)
&&e_{34} = (l_2 + l_5 + l_6 + l_7 + l_8)J_{34}\nonumber\\
&&~~~~~= (-l_1 - 2l_3 + 1)J_{34} = (2k_1 + 2k_2 - 1)J_{34}.%(-k_1 - k_3 + 1)
\label{eq17}
\end{eqnarray}

The ``turtle'' cluster contains only a half of the 20th-neighbor
pairs (see Fig.~15); therefore, we denote their contribution to
the energy density by $e_{20a}$.

We thus see that contributions $e_i$ to the energy density of all
the pairwise interactions  within the ``turtle'' cluster except
for the 21th neighbors, depend on $k_1$ and $k_2$. However, the
contribution of the 21th neighbors depends not only on these
quantities but also on $l_5$. Hence, only the pairwise interaction
of 21th neighbors or many-spin interactions, which includes 21th
neighbors, can give rise to a full-dimensional structure that
cannot be produced by any interaction within the ``windmill''
cluster. As we have already shown, this is the 1/7-plateau
structure (Fig.~13).

Having directly calculated the contributions of various pairwise
interactions, that do not enter the ``turtle'' cluster,  to the
energy density (up to the 43th neighbors) for 1/3-, 1/4-, 1/5-,
and 1/7-plateau structures as well as for the N\'{e}el structure,
we find that
\begin{eqnarray}
&&e_{19} = (k_1 + 2k_2 - 1)J_{19}\nonumber\\
&&e_{20} = 2e_{20a} = 2(2k_1 + 4k_2 - 2)J_{20},\nonumber\\
&&e_{23} = (-8k_1 - 8k_2 + 6)J_{23},\nonumber\\
&&e_{24} = (9k_1 + 6k_2 + 2l_5 - 5)J_{24},\nonumber\\
&&e_{27} = (11k_1 + 8k_2 + 2l_5 - 7)J_{27},\nonumber\\
&&e_{28} = (-4k_1 - 4k_2 + 3)J_{28},\nonumber\\
&&e_{29} = (6k_1 - 2)J_{29},\nonumber\\
&&e_{30} = (k_1 + 2k_2 - 1)J_{30},\nonumber\\
&&e_{31} = (-10k_1 - 8k_2 + 6)J_{31},\nonumber\\
&&e_{32} = (-16k_1 - 12k_2 - 8l_5 - \frac12 q + 10)J_{32},\nonumber\\%(-10k_1 - 8k_2 - 4l_5 - \frac12 q - 6)J_{32},\nonumber\\
&&e_{33} = (8k_1 + 8k_2 - 6)J_{33},\nonumber\\
&&e_{35} = (8k_1 + 6k_2 + 2l_5 - 5)J_{35},\nonumber\\
&&e_{36} = (18k_1 + 12k_2 + 4l_5 - 10)J_{36},\nonumber\\
&&e_{37} = (4k_1 + 4k_2 - 3)J_{37},\nonumber\\
&&e_{38} = (-22k_1 - 16k_2 + 14)J_{38},\nonumber\\
&&e_{39} = (-14k_1 - 12k_2 - 8l_5 - \frac12 q + 10)J_{39},\nonumber\\
&&e_{40} = (32k_1 + 24k_2 + 8l_5 - 20)J_{40},\nonumber\\
&&e_{41} = (9k_1 + 6k_2 + 2l_5 - \frac{11}{2})J_{41},\nonumber\\
&&e_{42} = (k_1 + 2k_2 - 1)J_{42},\nonumber\\
&&e_{43} = (-10k_1 - 8k_2 + 6)J_{43}.
\label{eq17a}
\end{eqnarray}
To prove these relations rigorously, one should consider clusters
larger than the ``turtle'' cluster but we do not do it here.

The contributions of the 32th-neighbor interaction to the energy
density for the 1/3-, 1/4-, 1/5-, and 1/7-plateau structures as
well as for the N\'{e}el structure are given by
\begin{eqnarray}
&&e_{32, N} = -2J_{32},~~ e_{32, 1/3} = -\frac23 J_{32},~~ e_{32, 1/4} = -J_{32},\nonumber\\
&&e_{32, 1/5} = -\frac25 J_{32},~~ e_{32, 1/7} = \frac27 J_{32}.
\label{eq17b}
\end{eqnarray}
The quantity $e_{32}$ (as well as $e_{39}$) cannot be written in
the form $ak_1 + bk_2 + cl_5 + d$, as one can do for all the
interactions up to the 31th neighbors and also for 33-38th and
40-43th neighbors. In addition to $k_1$, $k_2$, and $l_5$ a new
quantity, $q$, should be introduced. Thus, the 32th neighbor
interaction (as well as the 39th neighbors interaction) gives rise
to a new full-dimensional phase. This phase just corresponds to
the 1/4-plateau structure $(3,4)_{1,2}$ (Fig.~14).

The contributions of the 44th-neighbor interaction to the energy
density for the 1/3-, 1/4-, 1/5-, and 1/7-plateau structures as
well as for the N\'{e}el structure are given by
\begin{eqnarray}
&&e_{44, N} = J_{44},~~~~~~~~ e_{44, 1/3} = -\frac13 J_{44},\nonumber\\
&&e_{44, 1/4} = \frac12 J_{44},~~~~ e_{44, 1/5} = \frac15 J_{44},\nonumber\\
&&e_{44, 1/7} = -\frac17 J_{44},~~ e_{44, 1/9} = -\frac19 J_{44}.
\label{eq17c}
\end{eqnarray}
The 44th-neighbor interaction again gives rise to a new
full-dimensional phase, $(3,4)_4$, that is the 1/9-plateau phase.

Now we can conclude that new full-dimensional phases can emerge
from the boundary between phases 3 and 4 when new pairs of chains
begin to interact. Thus, the interactions of chains at the
distances of three and five square lattice constants (the seventh
and 21th neighbors) can give rise to the 1/5-, and 1/7-plateau
phases, respectively. The $1/n$-plateau phase, where $n$ is odd
(even) number, can emerge only provided the chains at the distance
of $n-2$ ($2n-2$) square lattice constants interact.

New plateaus can emerge in the following succession with
increasing interaction range: 1/5 plateau (the 7th neighbor
interaction, i.e, the interaction of chains at the distance of
three square lattice constants); 1/7 plateau (21th neighbors, 5
constants); 1/4 plateau (32th neighbors, 6 constants); 1/9 plateau
(44th neighbors, 7 constants); 1/5 plateau (8 constants); 1/11 and
3/11 plateaus (9 constants); 1/6 plateau (10 constants); 1/13 and
3/13 plateaus (11 constants); ... The 1/5 plateau which follows
the 1/9 plateau corresponds to a structure that differs from the
1/5-plateau structure shown in Fig.~9.

What sequences of phases (i.e., sequences of plateaus) are
possible with the field increase, if the interactions do not
overstep, for instance, the bounds of the ``turtle'' cluster?
There are four paths from the vertex $B$ to the vertex $A$ of the
tetrahedron $ABCD$ along its edges: $BA$, $BCA$, $BDA$, and
$BDCA$. These paths correspond to the sequences of plateaus
0--1/3, 0--1/5--1/3, 0--1/7--1/3, and 0--1/5--1/7--1/3. It depends
on the signs and values of interactions, which of these sequences
is realized.

If the interactions up to the 44th neighbors are involved into
consideration, then only the 1/9-, 1/7-, 1/5-, and 1/4-plateau
phases can emerge at the boundary between the N\'{e}el phase and
the 1/3-plateau phase. It depends on the interaction constants
$J_i$, which of these phases ``survive.'' The interaction
constants are determined by the exchange interaction between the
nearest neighbors and by the Ruderman-Kittel-Kasuya-Yosida (RKKY)
interaction.

The RKKY interaction in two dimensions reads \cite{bib22}
\begin{eqnarray}
&&H_{RKKY} = A^2\frac{m(k_F)^2}{8\pi}[J_0 (k_F R_{ij}) Y_0 (k_F R_{ij})\nonumber\\
&&~~~~~~~~~~~+ J_1 (k_F R_{ij}) Y_1 (k_F R_{ij})]\sigma_i\sigma_j,
\label{eq}
\end{eqnarray}
where $J_0$ and $J_1$ are the Bessel functions of the first kind
of the zero and first orders, respectively, $Y_0$ and $Y_1$ are
the Bessel functions of the second kind of the zero and first
orders, respectively, $k_F$ is the Fermi level, $R_{ij}$ is the
distance between the sites $i$ and $j$, $\sigma_i$ and $\sigma_j$
are the values of Ising spins at the sites $i$ and $j$, $m$ is an
effective mass of conduction electrons, and $A$ is the exchange
coupling constant.

It depends on the value of $k_F$, which of the 1/9-, 1/7-,\\ 1/5-,
and 1/4-plateau structures become full-dimensional, with the value
of $A$ influencing only the widths of the plateaus. In TmB$_4$ and
ErB$_4$, magnetic atoms of each layer are arranged in the
Archimedean lattice $3^2.4.3.4$ (see Fig.~16) which is
topologically equivalent to the SS lattice. For these compounds,
$\tilde J_1 = \tilde J_2$.

If, for instance, the nearest-neighbor interaction is
antiferromagnetic and equal to 1 ($\tilde J_1 = \tilde J_2 = 1$),
and the rest of $J_i$ ($i$ = 3--44) are determined by $k_F =
4.14/a$, where $a$ is the side of a square or a triangle on the
Archimedean lattice, then, among all the structures at the
boundary between phases 3 and 4, only the 1/9- and 1/7-plateau
structures become full-dimensional. Such plateaus are observed in
TmB$_4$, though in Ref.~\onlinecite{bib14} slightly different
structures are presented for these plateaus.

\begin{figure}[]
\begin{center}
\includegraphics[scale = 1.45]{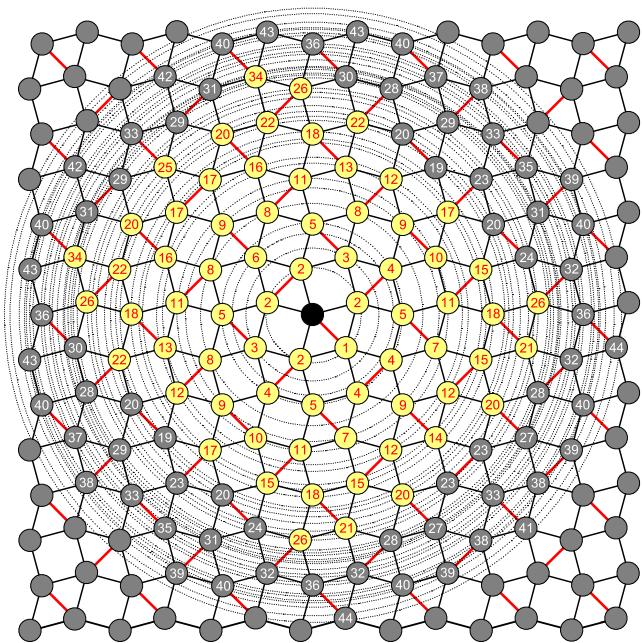}
\caption{Coordination circles and respective neighbors of the site
depicted in black on the Archimedean lattice $3^2.4.3.4$ that is
topologically equivalent to the SS lattice. The sites are
enumerated in the same manner as in Fig.~15. The sites which enter
a ``turtle'' cluster (see Fig.~4) together with the central
(black) site are depicted in yellow.}
\label{fig16}
\end{center}
\end{figure}

\section{Full-dimensional ground-state structures emerging from
the boundary between the 1/3-plateau and 1/2-plateau phases}

In a manner similar to the previous section, let us find the
ground-state structures that can emerge from the boundary between
the 1/3-plateau phase (phase 4) and the 1/2-plateau phase (phase
6) when the interaction range increases. The ground-state
structures at this boundary consist of the square configurations
\usebox{\suddu}, \usebox{\sduuu}, \usebox{\suduu}, \usebox{\suuuu};
\usebox{\uddu}, and \usebox{\duuu} \cite{bib13}. We consider the
configurations of the ``windmill'' cluster which generate all the
ground-state structures at the boundary between phases 4 and 6.
These are shown in Fig.~17. Configurations 1-3 and 12-21 generate
the structures of phase 6 (that is disordered itself), and
configurations 4 and 5 give rise to phase 4; configurations 6-11
are additional ones at the boundary between phases 4 and 6.

\begin{figure}[]
\begin{center}
\includegraphics[scale = 1.2]{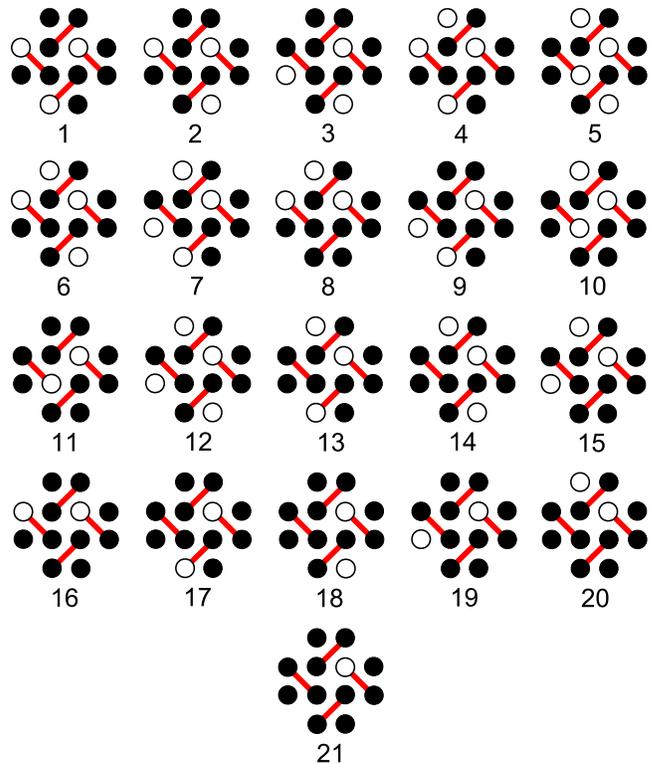}
\caption{Configurations of the ``windmill'' cluster for the
structures at the boundary between phases 4 and 6.}
\label{fig17}
\end{center}
\end{figure}

\begin{figure}[]
\begin{center}
\includegraphics[scale = 2.0]{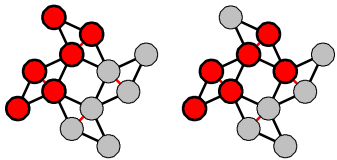}
\caption{``Seahorse'' (in red) is the maximum subcluster that can
occupy two nonequivalent positions in the ``windmill'' cluster.}
\label{fig18}
\end{center}
\end{figure}

\begin{figure}[]
\begin{center}
\includegraphics[scale = 1.5]{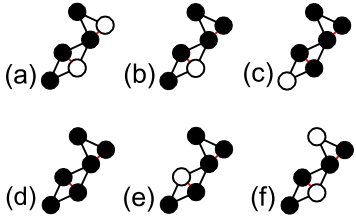}
\caption{Some configurations of the ``seahorse'' subcluster.}
\label{fig19}
\end{center}
\end{figure}

It should be noted that in this section we use the same notations
as in the previous one but here they denote other quantities. The
enumeration of the cluster configurations is also different.

Consider some configurations of the maximum subcluster which can
occupy two nonequivalent positions in the ``windmill'' cluster
(Fig.~18). These configurations are shown in Fig.~19. We put the
number of configurations (a) and (b) of the subcluster in
different positions equal and thus obtain the relations between
the fractional contents $k_i$ of the corresponding configurations
in the structures, i.e.,
\begin{eqnarray}
&&k_7 + k_9 + k_{13} + k_{17} = 0,\nonumber\\
&&k_7 - k_8 + k_9  - k_{10} - 2k_{11} + k_{13} - k_{15}\nonumber\\
&&- k_{16} + k_{17}- k_{19} - k_{20} - k_{21} = 0.
\label{eq18}
\end{eqnarray}
All the values of $k_i$ in these relations are equal to zero
(since $k_i$ should be nonnegative). Configurations (c) and (d)
lead to the relations that yield $k_{12}=0$ and $k_{18}=0$.
Configurations (e) and (f) lead to the relations
\begin{eqnarray}
&&k_1 - k_3 - k_6  - k_{14} = 0,\nonumber\\
&&k_1 - k_3 + k_4 - 2k_5 - k_{14} = 0.
\label{eq19}
\end{eqnarray}

The fact that $k_i = 0$ does not mean that the $i$th configuration
cannot enter the structures (all the configurations shown in
Fig.~17 can enter the structures at the boundary between phases 4
and 6). This only means that the number of such configurations is
infinitesimal as compared to the number of other configurations.

The fact that the major portion of the values of $k_i$ are equal
to zero can be proved by geometrical arguments. The fragment shown
in blue in Fig.~20 generates an infinite half-stripe composed with
configurations 1 and/or 4. If a configuration has two or three
such fragments, then these generate the relevant number of
half-stripes. For instance, configuration 15 gives rise to one
half-stripe, configuration 16 generates two perpendicular
half-stripes, configuration 17 also generates two half-stripes but
in opposite directions, and configuration 21 even gives rise to
three half-stripes (Fig.~21). Furthermore, and this is very
important, different copies of these configurations on the lattice
generate different copies of half-stripes. Just for this reason
the number of such configurations is infinitesimal compared to the
number of the rest of configurations. Similar geometrical
reasoning leads to the conclusion that $k_{14} = 0$, though the
latter does not follow from the relations between $k_i$.
\begin{figure}[]
\begin{center}
\includegraphics[scale = 1.2]{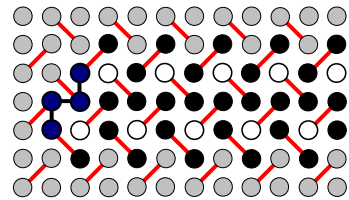}
\caption{A fragment (shown in blue) which generates a half-stripe
of configurations 1 and/or 4.}
\label{fig20}
\end{center}
\end{figure}

\begin{figure}[bth]
\begin{center}
\includegraphics[scale = 1.0]{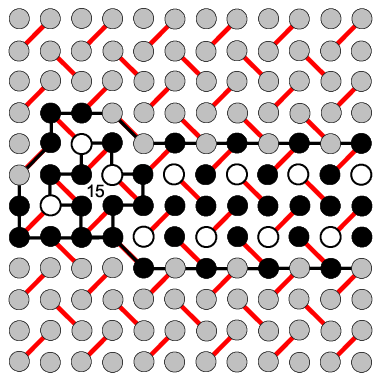}
\hspace{0.25cm}
\includegraphics[scale = 1.0]{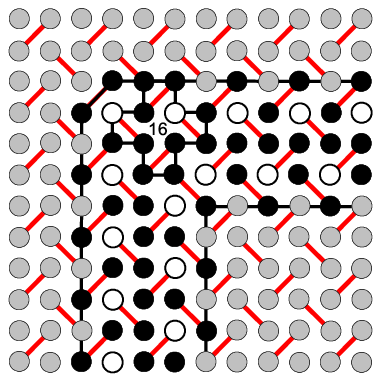}

\vspace{0.25cm}

\includegraphics[scale = 1.0]{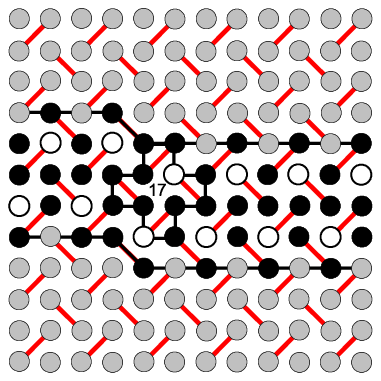}
\hspace{0.25cm}
\includegraphics[scale = 1.0]{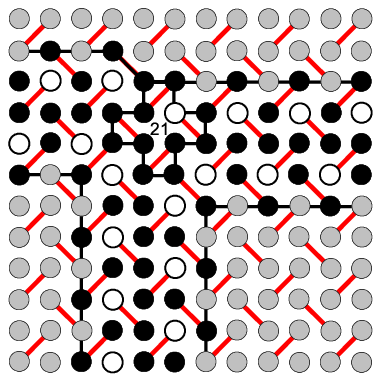}
\caption{Half-stripes generated by configurations 15, 16, 17, and
21.}
\label{fig21}
\end{center}
\end{figure}

Thus, only six initial configurations have nonzero fractional
contents in the ground-state structures on the boundary between
phases 4 and 6. These contents satisfy the relations given by
\begin{eqnarray}
&&k_1 = k_3 + k_6,\nonumber\\
&&k_2 = -2k_3 - 3k_5 - k_6 + 1,\nonumber\\
&&k_4 =  2k_5 - k_6.
\label{eq20}
\end{eqnarray}
We can exclude the configurations with $k_i = 0$ since their
number is infinitesimal. The remaining configurations generate
structures of the type shown in Fig.~22. These structures
represent a mixture of two kinds of stripes: one or two
antiferromagnetic chains bordered by ferromagnetic ones.

\begin{figure}[]
\begin{center}
\includegraphics[scale = 1.0]{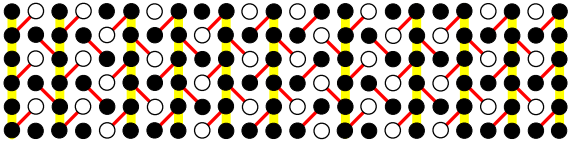}
\caption{An example of a structure at the boundary between phases
4 and 6.}
\label{fig22}
\end{center}
\end{figure}

The region $ABCD$ of variation for the quantities $k_3$, $k_5$,
and $k_6$ is shown in Fig.~23. The points $A$, $B$, and $C$
correspond to the structures $6a$, $6b$ (Fig.~24), and 4
[Fig.~4(b)], respectively. Structure $6a$ does not contain
configurations 3, 5, and 6; it is generated by configuration 2.
Structures $6b$ and 4 contain maximum possible numbers of
configurations 3 and 5, respectively. The structure with the
maximum possible number of configurations 6 corresponds to point
$D$. It is shown in Fig.~25. For $0 < k_5 < \frac17$ the maximum
of $k_6$ at $k_3=0$ is determined by the number of wide stripes
(as in structure 4), and for $\frac17 < k_5 < \frac13$ it is
determined by the number of narrow stripes (as in structures $6a$
and $6b$). Similarly, for nonzero $k_3$: the face $ABD$ ($k_6 =
2k_5$, $k_{3max}=\frac12 -\frac72 k_5$) corresponds to the
structures where each wide stripe contains configurations 6 on
both sides.

The fact that the region of variation for the quantities $k_3$,
$k_5$, and $k_6$ is the tetrahedron $ABCD$ shown in Fig.~23 can be
easily proved within the context of relations (19). Thus, the last
of these yields the inequality
\begin{equation}
k_6 = 2k_5 - k_4 \leqslant 2k_5,
\label{eq21}
\end{equation}
which, being taken for an equation, represents the equation of the
face $ABD$ (then $k_4 = 0$). The second of relations (19) and the
inequality $k_6 - k_2 \leqslant 0$ yield the inequality
\begin{equation}
2k_3 + 3k_5 + 2k_6 - 1 = k_6 - k_2 \leqslant 0,
\label{eq22}
\end{equation}
which, being taken for an equation, represents the equation of the
face $BCD$ (then $k_2 = k_6$). To complete the proof one should
show that some structures correspond to the points $A$, $B$, $C$,
and $D$. This was done before.

\begin{figure}[bht]
\begin{center}
\includegraphics[scale = 0.75]{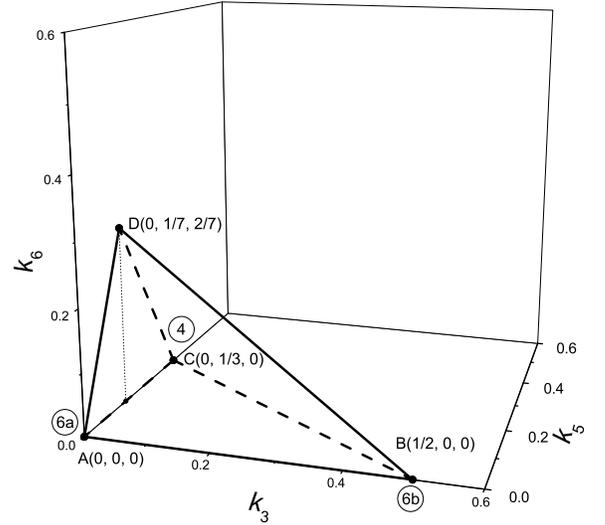}
\caption{Tetrahedron of variation for the quantities  $k_3$,
$k_5$, and $k_6$.}
\label{fig23}
\end{center}
\end{figure}

Let us find the magnetization and the contributions to the energy
density for the structures consisting of the configurations
considered. The magnetization per site reads
\begin{eqnarray}
&&m = \frac{1}{12}(6k_1 + 6k_2 + 6k_3 + 4k_4 + 4k_5 + 4k_6)\nonumber\\
&&~~~~= -\frac12 k_5 + \frac12,
\label{eq23}
\end{eqnarray}
and the contributions of pairwise interactions to the energy
density
\begin{eqnarray}
&&e_1 = \frac14(-k_1 + k_3 - k_4)\tilde J_1 = -\frac12 k_5\tilde J_1,\nonumber\\
&&e_2 = -2k_5\tilde J_2,\nonumber\\
&&e_3 = k_5J_3,\nonumber\\
&&e_4 = (2k_3 + 2k_5)J_4,\nonumber\\
&&e_5 = (k_1 + 2k_2 + k_3 + k_4 + 2k_6)J_5\nonumber\\
&&~~~~= (-2k_3 - 4k_5 + 2)J_5,\nonumber\\
&&e_6 = \frac14(k_1 - k_3 + 2k_4 + 2k_5 + k_6)J_6 = \frac32 k_5J_6,\nonumber\\
&&e_7 = (-k_1 + k_3 - k_4 + k_5)J_7 = -k_5 J_7,\nonumber\\
&&e_8 = (2k_1 - 2k_5 - 2k_6)J_8 = (2k_3 - 2k_5)J_8,\nonumber\\
&&e_{9a} = \frac12(2k_2 - k_4 + k_6)J_9 = \frac12(-4k_3 - 8k_5 + 2)J_9,\nonumber\\
&&e_{11} = (k_4 - k_6)J_{11} = (2k_5 - 2k_6)J_{11},\nonumber\\
&&e_{16} = (k_1 - k_3 + k_5 + k_6)J_{16} = (k_5 + 2k_6)J_{16}.
\label{eq24}
\end{eqnarray}

\begin{figure}[]
\begin{center}
\includegraphics[scale = 1.0]{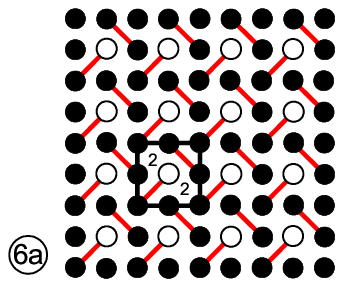}
\hspace{0.25cm}
\includegraphics[scale = 1.0]{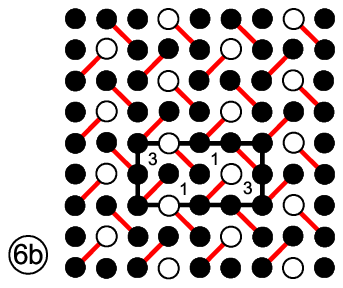}
\caption{Structures $6a$ and $6b$. They are stabilized by ferro-
and antiferromagnetic interactions of the fourth neighbors
(antiferro- and ferromagnetic interactions of the fifth
neighbors), respectively.}
\label{fig24}
\end{center}
\end{figure}

As we can follows from the above relations, the quantities $e_1$,
$e_2$, and $e_3$ depend on $k_5$ only (i.e., on the magnetization
$m$). Therefore, if there are no interactions other than $\tilde
J_1$, $\tilde J_2$, and $J_3$, then the set of configurations of
the ``windmill'' cluster (Fig.~17) generates only two
full-dimensional phases: phase 4 with the maximum number of
configurations 5, and phase 6 without configurations 5. The
quantities $e_6$ and $e_7$ also depend on $k_5$ only, therefore
the corresponding interactions do not lift the degeneracy at the
boundary between phases 4 and 6, nor in the phase 6 itself. The
quantities $e_4$, $e_5$, and $e_8$ depend not only on $k_5$ but
also on $k_3$. Each of the corresponding interactions lifts the
degeneracy of phase 6, giving rise to two new full-dimensional
phases: phase $6a$ and phase $6b$ (Fig.~24). The quantities
$e_{11}$ and $e_{16}$ depend on $k_1$ and $k_6$, hence the phase
shown in Fig.~25 is given rise by the interactions $J_{11}$ and
$J_{16}$.

The term with $k_3$ enters the expressions for $e_4$ and $e_8$
with the ``plus'' sign and the ones for $e_5$ --- with the
``minus'' sign. Thus, the phase $6a$ is stabilized by the
ferromagnetic interactions $J_4$ and $J_8$ as well as by the
antiferromagnetic interaction $J_5$ (and vice versa for the phase
$6b$). The term with $k_6$ enters the expressions for $e_{11}$
with the ``minus'' sign and in the expression for $e_{16}$ with
the ``plus'' sign; therefore, the phase shown in Fig.~25 is
stabilized by the antiferromagnetic interaction $J_{11}$ and by
the ferromagnetic interaction $J_{16}$.

\begin{figure}[]
\begin{center}
\includegraphics[scale = 1.0]{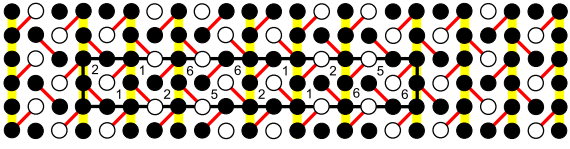}
\caption{Structure with the maximum value of $k_6 = \frac27$ (the
magnetization $m = \frac37$). It emerges from the boundary between
phases 4 and 6 when the interaction range reaches the 11th
neighbors (three square lattice constants).}
\label{fig25}
\end{center}
\end{figure}

\begin{figure}[bht]
\begin{center}
\includegraphics[scale = 1.0]{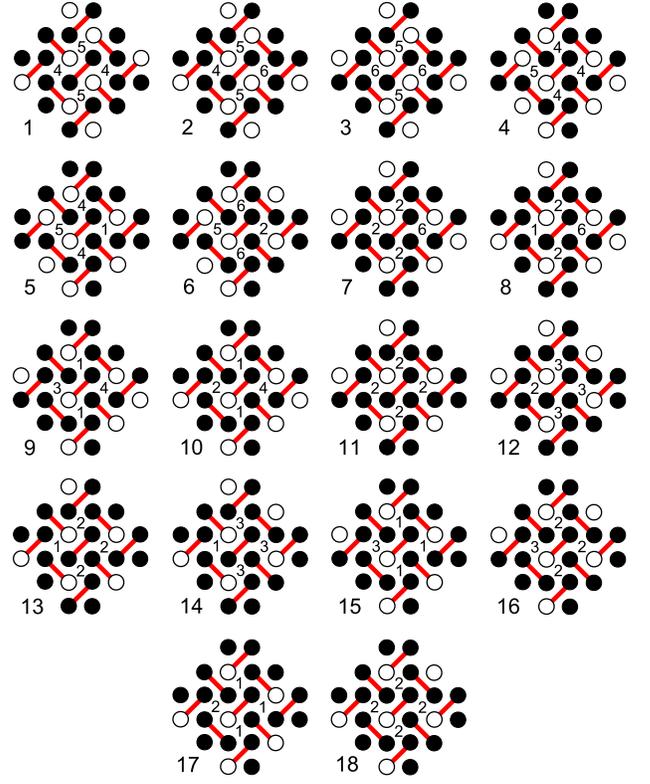}
\caption{Configurations of the ``turtle'' cluster generating all
the structures at the boundary between phases 4 and 6. In each
``empty'' square, a number of the ``windmill'' configuration with
the center in this square is indicated.}
\label{fig26}
\end{center}
\end{figure}

To investigate the effect of longer-range interactions, we
consider the configurations of the ``turtle'' cluster which
generate all the structures at the boundary of phases 4 and 6.
These configurations are depicted in Fig.~26. The maximum
subcluster which can occupy two nonequivalent positions in the
``turtle'' cluster (Fig.~11) generates a set of relations for the
corresponding fractional contents $l_i$ that is given by
\begin{eqnarray}
&&2l_1+l_2-l_4-l_5=0,\nonumber\\
&&l_2+2l_3-l_6=0,\nonumber\\
&&l_5-l_9-l_{10}=0,\nonumber\\
&&l_6-l_7-l_8=0,\nonumber\\
&&l_7+l_{11}-l_{18}=0,\nonumber\\
%&&l_7+2l_{11}+l_{13}-l_{16}-2l_{18}=0,\nonumber\\
&&l_{11}+l_{13}-l_{16}-l_{18}=0,\nonumber\\
&&l_8-l_{10}+l_{13}-l_{17}=0,\nonumber\\
&&l_9-l_{14}+l_{15}=0,\nonumber\\
&&l_{12}-l_{16}=0,\nonumber\\
&&\sum\limits_{i=1}^{18} l_i = 1.
\label{eq25}
\end{eqnarray}
The last relation is the normalization condition.

This set of equations yields
\begin{eqnarray}
&&l_2 = -3l_1+l_3-4l_{10}-2l_{14}-4l_{17}-2l_{18}+1,\nonumber\\
&&l_4 = -l_1+l_3-5l_{10}-3l_{14}+l_{15}-4l_{17}-2l_{18}+1,\nonumber\\
&&l_5 = l_{10}+l_{14}-l_{15},\nonumber\\
&&l_6 = -3l_1+3l_3-4l_{10}-2l_{14}-4l_{17}-2l_{18}+1,\nonumber\\
&&l_7 = -3l_1+3l_3-5l_{10}+l_{13}-2l_{14}-5l_{17}-2l_{18}+1,\nonumber\\
&&l_8 = l_{10}-l_{13}+l_{17},\nonumber\\
&&l_9 = l_{14}-l_{15},\nonumber\\
&&l_{11} = 3l_1-3l_3+5l_{10}-l_{13}+2l_{14}+5l_{17}+3l_{18}-1,\nonumber\\
&&l_{12} = 3l_1-3l_3+5l_{10}+2l_{14}+5l_{17}+2l_{18}-1,\nonumber\\
&&l_{16} = 3l_1-3l_3+5l_{10}+2l_{14}+5l_{17}+2l_{18}-1.
\label{eq26}
\end{eqnarray}
Hence, only eight of 18 quantities $l_i$ are independent.

It is easy to find the relations between the quantities $k_i$
($i=1-6$) and $l_i$ ($i=1-18$). We just have to calculate the
numbers of configurations of the ``windmill'' subcluster in
configurations of the ``turtle'' cluster. Thus we have
\begin{eqnarray}
&&4k_1=l_5+l_8+2l_9+2l_{10}+l_{13}+l_{14}+3l_{15}+3l_{17},\nonumber\\
&&~~~~~=4(l_{10}+l_{14}+l_{17}),\nonumber\\
&&4k_2=l_6+3l_7+2l_8+l_{10}+4l_{11}+l_{12}+3l_{13}\nonumber\\
&&~~~+3l_{16}+l_{17}+4l_{18},\nonumber\\
&&4k_3=l_9+3l_{12}+3l_{14}+l_{15}+l_{16}\nonumber\\
&&~~~~=4(3m - 1)-4(l_{10}+l_{17}+l_{18}),\nonumber\\
&&4k_4=2l_1+l_2+3l_4+2l_5+l_9+l_{10},\nonumber\\
&&4k_5=2l_1+2l_2+2l_3+l_4+l_5+l_6,\nonumber\\
&&4k_6=l_2+2l_3+2l_6+l_7+l_8.
\label{eq27}
\end{eqnarray}

Within the context of these and previous relations, we can write
the contributions into the energy density for the pairwise
interactions which (except for $J_{11}$ and $J_{16}$) are not
contained in the ``windmill'' cluster, i.e.,
\begin{eqnarray}
&&e_9 = (-4k_3 - 8k_5 + 2)J_9,\nonumber\\
&&e_{10} = k_5J_{10},\nonumber\\
&&e_{11} = (2k_5 - 2k_6)J_{11},\nonumber\\
&&e_{12} = (2k_3 + 6k_5 - 2k_6)J_{12},\nonumber\\
&&e_{13} = (-k_5 + 2k_6)J_{13},\nonumber\\
&&e_{14} = -\frac12 k_5J_{14},\nonumber\\
&&e_{15} = (4k_3 - 2k_5 + 4k_6 - 4l_{14})J_{15},\nonumber\\
&&e_{16} = (k_5 + 2k_6)J_{16},\nonumber\\
&&e_{17} = ( 2k_3 + 2k_5 - 2k_6)J_{17},\nonumber\\
&&e_{18} = (-4k_3 - 4k_5 - 2k_6 + 4l_{14} + 2)J_{18},\nonumber\\
&&e_{21} = (k_5 - k_6 + l_3)J_{21},\nonumber\\
&&e_{25} = (-\frac12 k_5 + 2k_6)J_{25}.
\label{eq28}
\end{eqnarray}

Pairwise interactions up to the 14th neighbors as well as
many-spin interactions which correspond to the clusters that
include only neighbors up to 14th range cannot give rise to any
full-dimensional structure from the boundary between phases 4 and
6 except for the four structures mentioned above. The reason is
that the contributions to the energy density $e_i$ ($i = 1-14$)
depend on $k_3$, $k_5$, and $k_6$ only. On the contrary, the
contributions $e_{15}$ and $e_{18}$ depend also on $l_{14}$;
therefore, the corresponding interactions can give rise to new
full-dimensional structures. It should be noted that the
15th-neighbor interaction is an interaction of chains at the
distance of four square lattice constants.

Since the structures at the boundary between phases 4 and 6 are
striped, the identical configurations are organized in stripes.
Therefore, a structure can be described by a sequence of
``windmill'' configurations.

Let us find new full-dimensional structures that can be given rise
by the 15th-neighbor interaction. The expression for $e_{15}$ can
be rewritten in terms of $k_1$, $k_5$, and $l_{14}$. The
polyhedron of variation for these quantities is shown in Fig.~27.
Vertices $A$, $B$, $C$, and $D$ correspond to the structures
similar to those in Fig.~23; vertex $E$ corresponds to the
structure shown in Fig.~28. Let us prove that the pyramid $ABCDE$
is indeed the region of variation for the quantities $k_1$, $k_5$,
and $l_{14}$ in the relevant space.

It is easy to show that $k_1$ and $k_5$ vary within the triangle
$ACM$. With this observation in view, we just have to prove that
the points with maximum (minimum) values of $l_{14}$, for fixed
$k_1$ and $k_5$, are the points of the triangle $ABC$ [of the
quadrangle $ACDE$ ($l_{14} = 0$) and  triangle $BDE$].

For fixed values of $k_1$ and $k_5$, the value of $l_{14}$ reaches
the maximum in the structures where each column of the
``windmill'' configurations 1 has two neighbor columns of
configurations 3 (if the number of configurations 5 makes it
possible); then $l_{14} = k_1$ (the face $ABC$ in Fig.~27).

Let the number of ``windmill'' configurations 1 and 5 be equal to
$N_1$ and $N_5$, respectively. Then the number of configurations 1
can exceed the number of configurations 3 by $2N_5$ at the most.
(If there are no configurations 5, then the number of
configurations 3 is equal to the number of configurations 1.) The
number of the rest $N_1 - 2N_5$ configurations 1 together with
similar number of configurations 3 and similar number of
configurations 2 is equal to $N - 7N_5$, where $N$ is the total
number of ``windmill'' configurations in the structure. If the
number of configurations 2 is at least twice greater than the
number of configurations 3, then configurations 1 and 3 can be
separated by configurations 2 so that no combinations 133 appear
(then $l_{14} = 0$). Then the number of configurations 1 cannot
exceed $2N_5 + \frac14 (N - 7N_5)$. Each excessive configuration 1
inevitably leads to the appearance of two ``turtle''
configurations 14. Hence, the minimum number of configurations 14
is given by
\begin{eqnarray}
&&l_{14} = 2\left[k_1 - \left(2k_5 +\frac14(1 -
7k_5)\right)\right]\nonumber\\
&&~~~= 2k_1 - \frac{k_5+1}{2}.
\label{eq29}
\end{eqnarray}
This is just the equation of the face $BDE$.

The equation of the face $ABC$ can be obtained in a similar
manner. The first of Eqs.~(20) can be rewritten in the form
\begin{equation}
l_{14} = k_1 - l_{10} - l_{17},
\label{eq30}
\end{equation}
whence it follows that
\begin{equation}
l_{14} \leqslant k_1.
\label{eq31}
\end{equation}
The relation that yields the equation of the face $ABC$ is
obtained under the condition $l_{10} = l_{17} = 0$.

The face $BDE$ gives the inequality
\begin{equation}
4k_1 - k_5 - 2l_{14} - 1 \leqslant 0.
\label{eq32}
\end{equation}
Using the relations for $l_i$ and $k_i$, this inequality can be
reduced to
\begin{equation}
2l_1 + l_2 + l_{18} \geqslant 0.
\label{eq33}
\end{equation}
This proves that the face $BDE$ was obtained correctly and in this
face $l_1 = l_2 = l_{18} = 0$. The relations for $l_i$ yield also
$l_4 = l_5 = l_7 = l_9 = l_{10} = l_{11} = 0$.

Hence, the 15th-neighbor anfiferromagnetic interaction gives rise
to the structure shown in Fig.~28. This structure has the number
of ferro- and antiferromagnetic SS dimers similar to the
1/2-plateau structure proposed in Ref.~\onlinecite{bib14}.
However, the dimensions of their unit cells are different: 8 and
16 square lattice constants, respectively; ferro- and
antiferromagnetic chains are also distributed differently.

\begin{figure}[]
\begin{center}
\includegraphics[scale = 0.75]{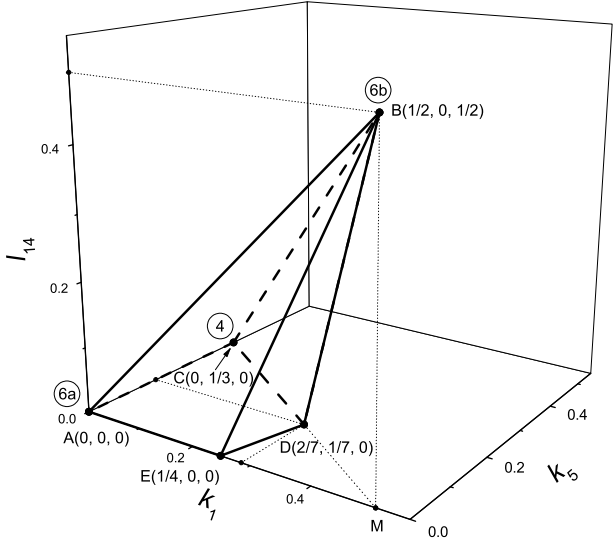}
\caption{Polyhedron of variation for the quantities $k_1$, $k_5$,
and $l_{14}$. Each vertex corresponds to a full-dimensional
structure.}
\label{fig27}
\end{center}
\end{figure}

\begin{figure}[]
\begin{center}
\includegraphics[scale = 1.0]{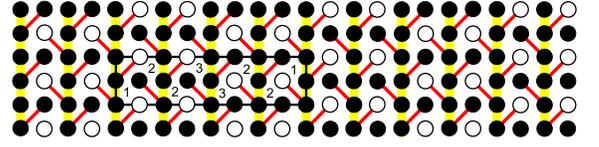}
\caption{The structure given rise by the antiferromagnetic
15th-neighbor interaction from the boundary between phases 4 and
6. This structure possesses maximum number of ``windmill''
configurations 1 (or 3) with no ``windmill'' configurations 5 and
``turtle'' configurations 14.}
\label{fig28}
\end{center}
\end{figure}

\section{Full-dimensional ground-state structure emerging from the
boundary between phases 1 and 6}

All the structures at the boundary between phases 1 and 6 consist
of the configurations of squares \usebox{\suddu}, \usebox{\sduuu},
\usebox{\suduu}, \usebox{\suuuu}; \usebox{\duuu}, and \usebox{\uuuu},
or with the set of configurations of the ``screw'' cluster shown
in Fig.~29. In addition to the normalization condition, there are
three relations for the fractional contents $p_i$ ($i=$1-10) of
these configurations. It is easy to derive them, by considering
the configurations of the maximum subcluster which can occupy two
nonequivalent positions in the ``screw'' cluster (Fig.~30).
\begin{eqnarray}
&&p_4 - p_7 + 2p_8 + 4p_{10}= 0,\nonumber\\
&&p_2 - p_3 + p_4 + 2p_5 + p_7 + p_8 + p_9 = 0,\nonumber\\
&&\sum\limits_{i=1}^{10} p_i = 1.
\label{eq34}
\end{eqnarray}

\begin{figure}[bth]
\begin{center}
\includegraphics[scale = 1.2]{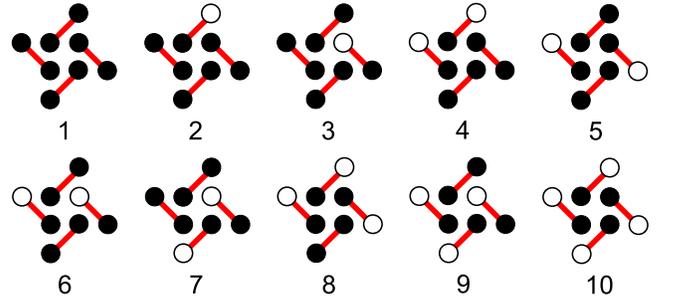}
\caption{Configurations of the ``screw'' cluster for the boundary
between phases 1 and 6.}
\label{fig29}
\end{center}
\end{figure}

\begin{figure}[bth]
\begin{center}
\includegraphics[scale = 1.2]{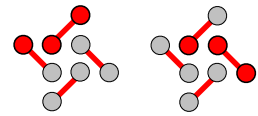}
\caption{The maximum subcluster that can occupy two nonequivalent
positions in the ``screw'' cluster.}
\label{fig30}
\end{center}
\end{figure}

Let us find the magnetization per site, taking into account that
two sites are the share of each of the ``screw'' clusters on the
lattice
\begin{eqnarray}
&&m = \frac12(2p_1 + 2p_2 + p_3 + 2p_4 + 2p_5 + p_6 + p_7 + 2p_8\nonumber\\
&&~~~~~+ p_9 + 2p_{10}) = \frac12 (2 - p_3 - p_6 - p_7 - p_9).
\label{eq35}
\end{eqnarray}

The contributions into the energy density of pairwise interactions
within the ``screw'' cluster are given by
\begin{eqnarray}
&&e_1 = \frac18(4p_1 + 2p_2 + 2p_3 - 2p_8 - 2p_9 - 4p_{10})\tilde J_1\nonumber\\
&&~~~~ = (m - \frac12)\tilde J_1,\nonumber\\
&&e_2 = \frac12(4p_1+4p_2+4p_4+4p_5+4p_8+4p_{10})\tilde J_2\nonumber\\
&&~~~~ = 2(2m-1)\tilde J_2,\nonumber\\
&&e_3 = \frac12(2p_1+2p_2+2p_4+2p_5+2p_8+2p_{10})J_3\nonumber\\
&&~~~~ = (2m-1)J_3,\nonumber\\
&&e_4 = \frac12(4p_1+2p_2+2p_3+4p_7-2p_8+2p_9-4p_{10})J_4\nonumber\\
&&~~~~ = 2(1-p_3-p_6)J_4,\nonumber\\
&&e_5 = \frac12(4p_1+2p_2+2p_3+4p_6-2p_8+2p_9-4p_{10})J_5\nonumber\\
&&~~~~ = 2(1-p_3-p_7)J_5,\nonumber\\
&&e_7 = (p_1+p_3-p_4+p_5-p_9+p_{10})J_7\nonumber\\
&&~~~~ = [4m-3+(p_3+p_6)+(p_3+p_7)\nonumber\\
&&~~~~ -(p_2+2p_4+p_8)]J_7.
\label{eq36}
\end{eqnarray}

\begin{figure}[]
\begin{center}
\includegraphics[scale = 0.75]{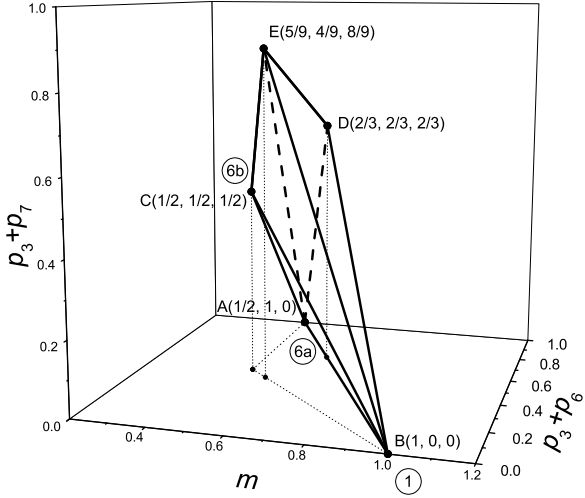}
\caption{Polyhedron of variation for the quantities $m$,
$p_3+p_6$, and $p_3 + p_7$.}
\label{fig31}
\end{center}
\end{figure}

\begin{figure}[h]
\begin{center}
\includegraphics[scale = 1.0]{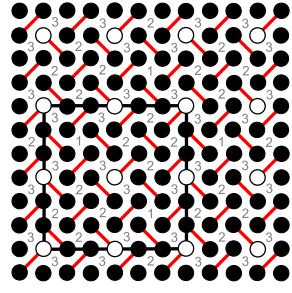}
\caption{A structure emerging from the boundary between phases 1
and 6. It corresponds to the vertex $D$ for which $p_3 = \frac12$,
$p_7 = \frac13$, $p_8 =\frac16$; and $m = \frac23$.}
\label{fig32}
\end{center}
\end{figure}

\begin{figure}[h]
\begin{center}
\includegraphics[scale = 0.75]{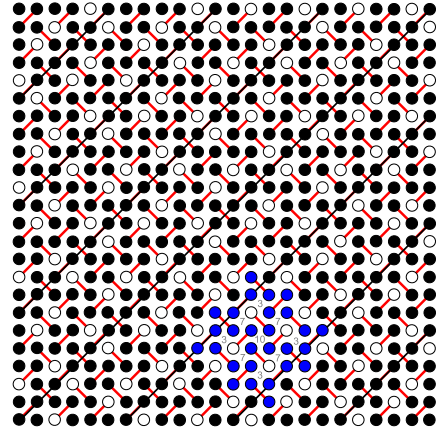}
\caption{Partially disordered structure emerging from the boundary
between phases 1 and 6. It corresponds to the vertex $E$ for which
$p_3 = p_7 = \frac49$, $p_{10} =\frac19$; and $m = \frac59$.}
\label{fig33}
\end{center}
\end{figure}

The region of variation for the quantities $m$, $p_3 + p_6$, and
$p_3 + p_7$ is the polyhedron $ABCDE$ shown in Fig.~31. It is
described by the set of inequalities
\begin{eqnarray}
&&ABC: 2m + (p_3 + p_6) + (p_3 + p_7) - 2 \geqslant 0,\nonumber\\
&&ABD: 2m + (p_3 + p_6) - 2 \leqslant 0,\nonumber\\
&&BCE: m + (p_3 + p_6) - 1 \geqslant 0,\nonumber\\
&&BDE: 2m + (p_3 + p_7) - 2 \leqslant 0,\nonumber\\
&&ACE: 6m - (p_3 + p_6) - (p_3 + p_7) - 2 \geqslant 0,\nonumber\\
&&ADE: 2m - 3(p_3 + p_6) - 2(p_3 + p_7) + 2 \geqslant 0.
\label{eq37}
\end{eqnarray}
These inequalities are rather difficult to find but easy to prove
using the expression for $m$ and the relations for $p_i$. For
instance, the inequality
\begin{equation}
2m + p_3 + p_6 - 2  \leqslant 0
\label{eq38}
\end{equation}
immediately follows from the expression (37) for $m$. It becomes
an equation if $p_7 = p_9 = 0$.

When proving the inequalities, we find that at faces some of
quantities $p_i$ are equal to zero. We have

\begin{eqnarray}
&&ABC: p_2 = p_4 = p_5 = p_7 = p_8 = 0,~~ p_{10}=0;\nonumber\\
&&ABD: p_7 = p_9 = 0;\nonumber\\
&&BCE: p_2 = p_4 = p_5 = p_6 = p_8 = 0;\nonumber\\
&&BDE: p_6 = p_9 = 0;\nonumber\\
&&ACE: p_1 = p_2 = p_4 = p_5 = p_8 = 0;\nonumber\\
&&ADE: p_1 = p_2 = p_4 = p_9 = 0.
\label{eq39}
\end{eqnarray}

The ground-state structures for the edges of the polyhedron
$ABCDE$ are constructed with the following configurations of the
``screw'' cluster:
\begin{eqnarray}
&&AB: 1, 3, 6;~AC: 3, 6, 9;~AD: 3, 5, 6;\nonumber\\
&&AE: 3, 6, 7, 10;~BC: 1, 3, 9;\nonumber\\
&&BD: 1, 2, 3, 4, 5, 8, 10;~BE: 1, 3, 7, 10;\nonumber\\
&&CE: 3, 7, 9, 10;~DE: 3, 5, 7, 8, 10.
\label{eq40}
\end{eqnarray}

At the vertex $E$, four faces converge: $BCE$, $BDE$, $ACE$, and
$ADE$, or edges $AE$, $BE$ , $CE$, and $DE$. Thus, the
ground-state structures in this vertex consist of configurations
3, 7, and 10 of the ``screw'' cluster.  For this vertex, Eqs. (33)
reduce to the set
\begin{eqnarray}
&&p_7 - 4p_{10} = 0, \nonumber\\
&&-p_3 + p_7 = 0, \nonumber\\
&&p_3 + p_7 + p_{10} = 1.
\label{eq41}
\end{eqnarray}
The solution of this set of equations is $p_3 = \frac49$, $p_7 =
\frac49$, $p_{10} = \frac19$. The structure which corresponds to
the vertex $E$ is shown in Fig.~32.

At the vortex $D$ three faces converge: $ABD$, $BDE$, and $ADE$,
or edges $AD$, $BD$, and $ED$. Thus, the ground-state structures
in this vertex consist of configurations 3 and 5 of the ``screw''
cluster. For this vertex, Eqs. (33) reduce to the set
\begin{eqnarray}
&&-p_3 + 2p_5 = 0, \nonumber\\
&&~~~p_3 + p_5 = 1.
\label{eq42}
\end{eqnarray}
The solution of this set of equations is $p_3 = \frac23$, $p_5 =
\frac13$. The structure which corresponds to the vertex $D$ is
shown in Fig.~33.

For $6a$ $p_6 = 1$, and for $6b$ $p_3 = p_9 = \frac12$. In the
face $ABC$ $p_3 = p_9$, $p_2 = p_4 = p_5 = p_8 = p_{10} = 0$.

\begin{eqnarray}
&&k_2-k_4+2k_5+k_7+k_8+k_9+k_{12}-k_{13}-k_{15}\nonumber\\
&&+k_{16}+k_{18}+k_{19}+k_{20}=0,\nonumber\\
&&k_8-k_{12}+2k_{16}+k_{18}+4k_{26}+k_{28}=0,\nonumber\\
&&k_3-k_4+2k_6+k_7+k_9+k_{10}-k_{11}-k_{12}-k_{14}\nonumber\\
&&+k_{15}+k_{17}+k_{18}+k_{19}-k_{20}+k_{22}=0,\nonumber\\
&&k_{10}-k_{15}+2k_{17}+k_{19}-k_{22}+4k_{27}=0,\nonumber\\
&&k_{13}-k_{14}+k_{21}+k_{23}-k_{25}+k_{28}=0.
\label{eq43}
\end{eqnarray}

\begin{figure}[]
\begin{center}
\includegraphics[scale = 1.2]{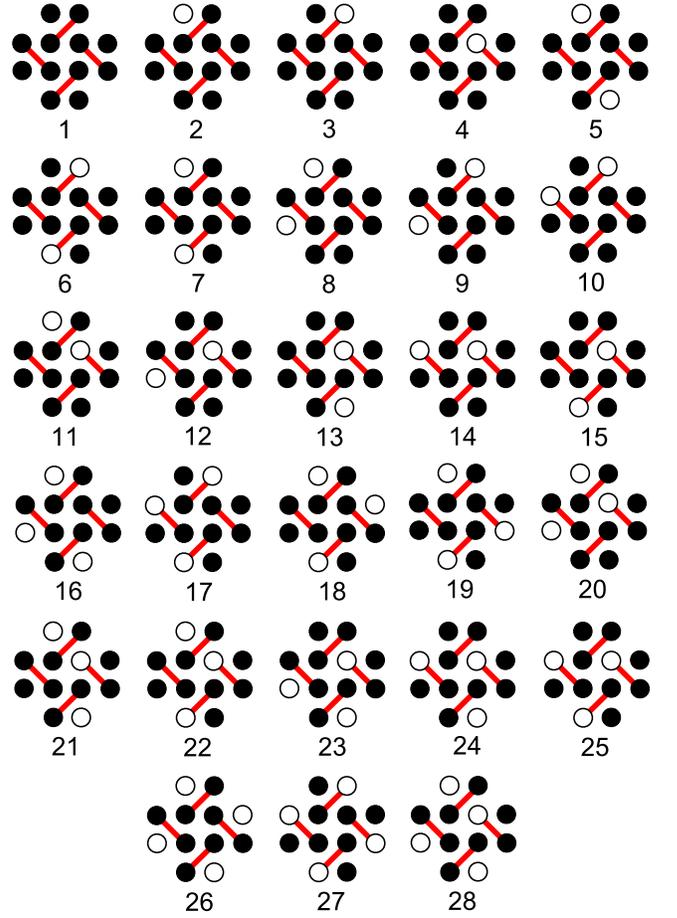}
\caption{Configurations of the ``windmill'' cluster for the
boundary between phases 1 and 6.}
\label{fig34}
\end{center}
\end{figure}

\section{Full-dimensional ground-state structures emerging from the
boundary between phases 3 and 7}

The 1/9-plateau structure and other structures of this type that
have been observed experimentally in Ref.~\cite{bib14} are the
ground-state structures at the boundary between phases 3 and 7.
Hence, it is interesting to investigate which full-dimensional
structures from this boundary are produced by the extended-range
interactions. All the structures at this boundary consist of two
configurations of the ``screw'' cluster (see Fig.~35) or with the
square configurations \usebox{\suddu}, \usebox{\sduud},
\usebox{\suduu}, \usebox{\suuuu}; \usebox{\uddu}, and
\usebox{\duuu}. Only one relation between fractional
contents $p_1$ and $p_2$ of ``screw'' configurations in structures
(the normalization condition) holds, i.e.,
\begin{equation}
p_1 + p_2 = 1.
\label{eq44}
\end{equation}

\begin{figure}[]
\begin{center}
\includegraphics[scale = 1.25]{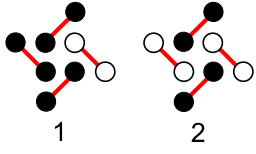}
\caption{Configurations of the ``screw'' cluster at the boundary
between phases 3 and 7.}
\label{fig35}
\end{center}
\end{figure}

\begin{figure}[]
\begin{center}
\includegraphics[scale = 1.25]{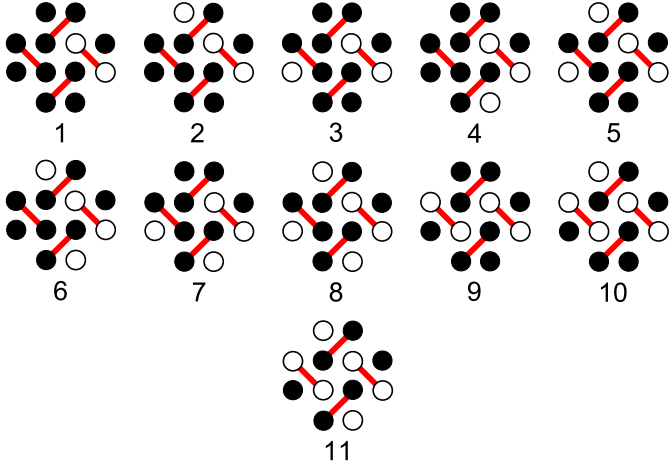}
\caption{Configurations of the ``windmill'' cluster for the
boundary between phases 3 and 7.}
\label{fig36}
\end{center}
\end{figure}

Now, let us consider a bigger cluster, the ``windmill'' cluster,
and its configurations generating all the ground-state structures
at the boundary between phases 3 and 7 (Fig.~36). An example of a
structure at this boundary is shown in Fig.~37. The number of the
``windmill'' configuration with the center in an "empty" square is
indicated in each of these. Considering configurations of the
``seahorse'' subcluster yields the relations between the
fractional contents $k_i$ of the ``windmill'' configurations in
structures, i.e.,
\begin{eqnarray}
&&k_8 = k_3,\nonumber\\
&&k_1-k_3+k_4-k_5 = 0,\nonumber\\
&&k_4+k_6+k_7+k_8-2k_9-k_{10} = 0.
\label{eq45}
\end{eqnarray}

\begin{figure}[htb]
\begin{center}
\includegraphics[scale = 1.0]{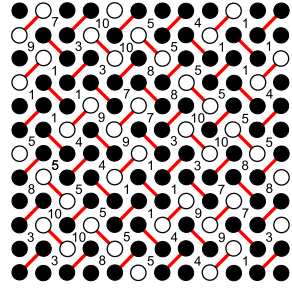}
\caption{An example of a structure at the boundary between phases
3 and 7. The number of the ``windmill'' configuration in an
``empty'' square with the center in this square is indicated in
each such square. $k_1 = k_5 =\frac15$, $k_3 = k_4 = k_7 = k_8 =
k_9 = k_{10} = \frac{1}{10}$.}
\label{fig37}
\end{center}
\end{figure}

It is easy to find relations between the quantities $p_1$, $p_2$,
and $k_i$. We have
\begin{eqnarray}
&&p_1 = k_1+k_2+k_3+k_4+k_5+k_6+k_7+k_8,\nonumber\\
&&p_2 = k_9+k_{10}+k_{11}.
\label{eq46}
\end{eqnarray}

The magnetization and the contributions into the energy density of
pairwise interactions within the ``windmill'' cluster are given by
\begin{eqnarray}
&&m = \frac{p_1}{2},\nonumber\\
&&e_1 = \frac{\tilde J_1}{2},\nonumber\\
&&e_2 = -2p_2\tilde J_2 = (4m - 2)\tilde J_2,\nonumber\\
&&e_3 = p_2J_3 = (-2m + 1)J_3,\nonumber\\
&&e_4 = -2p_2J_4 = (4m - 2)J_4,\nonumber\\
&&e_5 = 2p_2J_5 = (-4m + 2)J_5,\nonumber\\
&&e_6 = \frac14(k_1+2k_2+k_5+k_6-k_7+k_{10}+2k_{11})J_6\nonumber\\
&&~~~~= \frac12(-k_5-k_7-2k_8-2k_9-k_{10}+1)J_6\nonumber\\
&&~~~~= \frac12\left[-(k_5+k_7+2k_8)+(k_{10}+2k_{11})+4m-1\right]J_6,\nonumber\\
&&e_7 = p_2J_7 = (-2m + 1)J_7,\nonumber\\
&&e_8 = 2(k_1+2k_3+k_5-k_6+k_7-k_{10}-2k_{11})J_8\nonumber\\
&&~~~~= 2(k_5+k_7+2k_8-k_9-k_{10}-k_{11})J_8\nonumber\\
&&~~~~= 2(k_5+k_7+2k_8+2m-1)J_8,\nonumber\\
&&e_{9a} = \frac12(k_1+2k_2+k_5+k_6-k_7+k_{10}+2k_{11})J_9\nonumber\\
&&~~~~= \left[-(k_5+k_7+2k_8)+(k_{10}+2k_{11})+4m-1\right]J_9,\nonumber\\
&&e_{11} = (k_1+2k_3+k_5-k_6+k_7-k_{10}-2k_{11})J_{11}\nonumber\\
&&~~~~~ =  (k_5+k_7+2k_8+2m-1)J_{11},\nonumber\\
&&e_{16} = (k_1-k_5+k_6-k_7+k_9+k_{11})J_{16}\nonumber\\
&&~~~~~ =  (-2k_2-4k_5-2k_7-4k_8\nonumber\\
&&~~~~~~~~~ -3k_9-3k_{10}-k_{11}+2)J_{16}.
\label{eq47}
\end{eqnarray}

Structure 3 consists of configurations 11 and structure 7 of
configurations 1, 2, and 5. The 6th (and also 9ath) neighbor
interaction lifts the degeneracy of phase 7, the structures $7a$
and $7b$ (Fig.~38) thus become full-dimensional for $J_6 > 0$ and
$J_6 < 0$, respectively (and vice versa for the 8th neighbor
interaction ). For $7a$, we have $k_1 = k_5 = \frac12$. For $7b$,
we have $k_2 = 1$. For $J_6$ the expression attains its minimum
value for the structure $4a$ that consist of configurations 7 and
9 ($k_7 = \frac23$, $k_9 = \frac13$).

It follows from the inequality
\begin{equation}
2m = p_1 = k_1 + k_2 + ... + k_8 \geqslant k_5 + k_7 + 2k_8
\label{eq48}
\end{equation}
that the quantity $k_5 + k_7 + 2k_8$ cannot exceed $2m$. It is
equal to $2m$ for the structures formed by stripes which consist
of even numbers of antiferromagnetic chains between two
ferromagnetic ones.

\begin{figure}[tbh]
\begin{center}
\includegraphics[scale = 1.0]{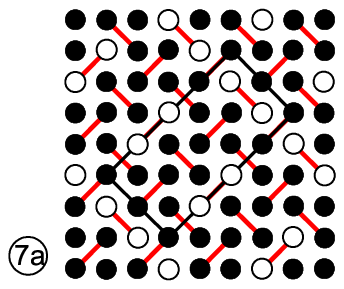}
\hspace{0.25cm}
\includegraphics[scale = 1.0]{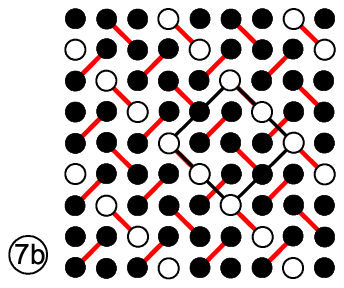}

\vspace{0.25cm}

\includegraphics[scale = 1.0]{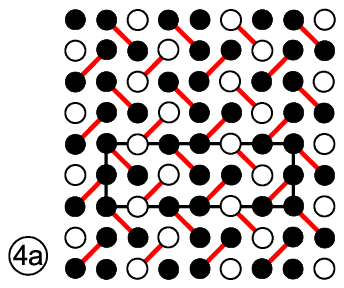}
\caption{Structures $7a$, $7b$, and $4a$.}
\label{fig38}
\end{center}
\end{figure}

\begin{figure}[]
\begin{center}
\includegraphics[scale = 0.75]{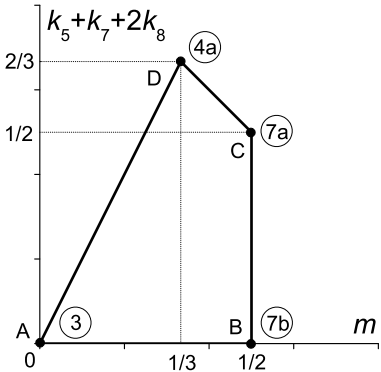}
\caption{Polygon of variation for the quantities $k_5+k_7+2k_8$,
and $m$. Its vertices correspond to the structures shown in
Fig.~37 and the N\'{e}el structure 3.}
\label{fig39}
\end{center}
\end{figure}

\begin{figure}[htb]
\begin{center}
\includegraphics[scale = 1.0]{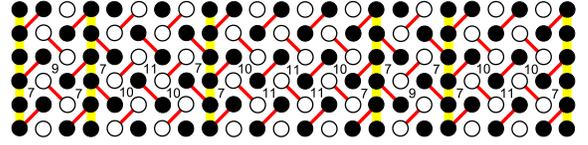}
\caption{An example of a structure at the boundary between phases
3 and 7. The structure of this kind of structures can be treated
as a simple mixture of structures 3 and $4a$. The number of the
``windmill'' configuration with the center in an "empty" square is
indicated in each square.}
\label{fig40}
\end{center}
\end{figure}

For ``pure'' structures, where all stripes are identical, the
magnetization and the values of $k_i$ are given by
\begin{eqnarray}
&&m = \frac{1}{2n+3},\nonumber\\
&&k_7 = k_{10} = \frac{2}{2n+3},~ k_{11} = \frac{2n-1}{2n+3},
\label{eq49}
\end{eqnarray}
where $2(n+1)$ $(n = 1, ...)$ are the numbers of antiferromagnetic
chains in the stripes. The structures of this type with fractional
values of magnetization $m = 1/7, 1/9, 1/11$ and some others have
been observed experimentally in TmB$_4$ \cite{bib14}. These
structures are generated by the set of square configurations
\usebox{\suddu}, \usebox{\sduud}, \usebox{\suduu}; \usebox{\uddu},
and \usebox{\duuu} (without configuration \usebox{\suuuu}), or the
equivalent set of configurations 7-11 of the ``windmill''
cluster.(There can be no more than two configurations 8 in a
structure; such a configuration emerges when an extreme
ferromagnetic chain forms a right angle.) These structures are
very similar to the structures at the boundary between phases 3
and 4, however, their antiferromagnetic chains are shifted.

\begin{figure}[htb]
\begin{center}
\includegraphics[scale = 1.0]{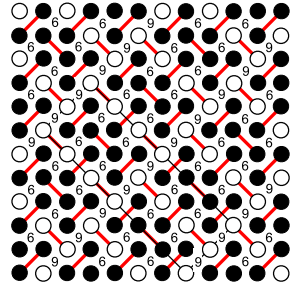}
\caption{The structure at the boundary between phases 3 and 7
which corresponds to the vertex $E$ of the polyhedron $ABCDE$.
Each ``empty'' square is labeled by the number of the ``windmill''
configuration with the center in this square. The structure is
generated by the ``windmill'' configurations 9 and 6 ($k_6 =
\frac23$, $k_{9} = \frac13$).}
\label{fig41}
\end{center}
\end{figure}

\begin{figure}[h]
\begin{center}
\includegraphics[scale = 0.75]{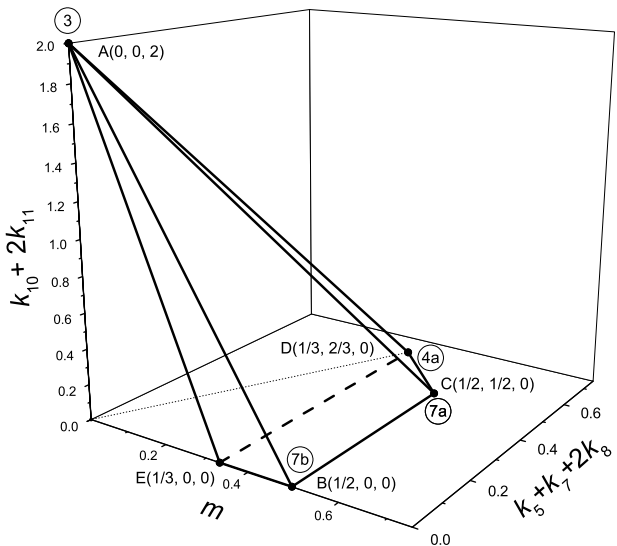}
\caption{Polyhedron of variation for the quantities $k_5+k_7+2k_8$
and $k_{10} + 2k_{11}$. Its vertices correspond to the structures
shown in Figs.~38 and 41 and the N\'{e}el structure 3.}
\label{fig42}
\end{center}
\end{figure}

The structure shown in Fig.~42 is given rise by the interactions
of sixth and eighth neighbors, though each of these interactions
separately cannot produce this structure. It mixes with structures
3, $7a$, and $7b$.

The expression for $m$ and the relations between $k_i$ yield the
equation
\begin{eqnarray}
&k_{10} + 2k_{11} = 2(1-2m)-2k_9-k_{10}\nonumber\\
&~~~~~~~~= 2(1-2m)-(k_4+k_6+k_7+k_8)\nonumber\\
&~~~~~~~~= 2(1-3m)+(k_1+k_2+k_3+k_5),
\label{eq50}
\end{eqnarray}
whence it follows that
\begin{equation}
2(1-2m) \leqslant k_{10} + 2k_{11} \leqslant 2(1-3m).
\label{eq51}
\end{equation}
Now we just have to prove that the following inequality holds:
\begin{equation}
m+(k_5+k_7+2k_8)+\frac12(k_{10} + 2k_{11}) - 1 \leqslant 0,
\label{eq52}
\end{equation}
which, when taken for an equation is the equation of face $ACD$.
Using the expression for $m$ and the relations between $k_i$, we
can transform this inequality into
\begin{equation}
k_2+k_4+2k_6  \geqslant 0,
\label{eq53}
\end{equation}
The latter holds because $k_i \geqslant 0$. Thus, we have proved
that the polyhedron $ABCDE$ (Fig.~42) is the region of variation
for the quantities $m$, $k_5+k_7+2k_8$, and $k_{10}+2k_{11}$. In
the face $ACD$, we have $k_2 = k_4 = k_6 = 0$.

\section{Conclusions}

We employ the relations for the fractional contents of the cluster
configurations in the structures generated by these configurations
to proposed a way to reveal the interactions which lift the
degeneracy at the boundary between the full-dimensional
ground-state phases and to construct the full-dimensional
structures which emerge consequently. We consider several
boundaries between full-dimensional ground-state phases for the
system of Ising spins on the Shastry-Sutherland lattice in a
magnetic field with the first-, second, and third-neighbor
interactions.

The seventh-neighbor interaction (one of the interactions between
chains at the distance of three square lattice constants on the SS
lattice) can partially lift the degeneracy between the N\'{e}el
phase and the 1/3-plateau phase giving rise to a full-dimensional
structure with the magnetization 1/5. The 21th neighbor
interaction (corresponding to the distance of five square lattice
constants) can generate a full-dimensional structure with the
magnetization 1/7. The 1/7 plateau as well as 1/9 and 1/11
plateaus were observed in TmB$_4$. The 1/9 and 1/11 plateaus can
emerge from the boundary between the N\'{e}el phase and the
1/3-plateau phase only provided the interaction between the chains
at the distances of seven and nine square lattice constants,
respectively, has begun.

\section{Acknowledgments}

The author is grateful to I. Stasyuk and T. Verkholyak for useful
discussions and suggestions and to O. Kocherga for correction of
the text.

\end{document}